\documentclass[twocolumn]{aastex62}
\submitjournal{ApJ}
\usepackage{enumitem}
\usepackage{graphicx,url}
\usepackage{color}

\def\gx339{GX~339$-$4}

\def\rxte{{\it RXTE}}
\def\xmm{{\it XMM-Newton}}
\def\suzaku{{\it Suzaku}}
\def\ngso{{\it Neil Gehrels Swift Observatory}}
\def\swift{{\it Swift}}
\def\nustar{{\it NuSTAR}}

\def\nicer{{\it NICER}}
\def\xillver{{\tt xillver}}

\def\xillverCp{{\tt xillverCp}}
\def\relxill{{\tt relxill}}

\def\relxillCp{{\tt relxillCp}}
\def\relxilllpCp{{\tt relxilllpCp}}

\def\nthComp{{\tt nthComp}}

\def\msun{$M_{\odot}$}

\def\redcsq{$\chi_{\nu}^2$}
\def\rin{$R_\mathrm{in}$}
\def\risco{$R_\mathrm{ISCO}$}
\def\ledd{$L_\mathrm{Edd}$}

\shorttitle{X-ray Reflection Spectroscopy of GX~339$-$4}
\shortauthors{Garc\'{\i}a \& et al.}

\begin{document}



\title{\large\bf The 2017 Failed Outburst of GX~339$-$4: Relativistic X-ray Reflection 
near the Black Hole Revealed by {\it NuSTAR} and {\it Swift} Spectroscopy}

\correspondingauthor{Javier~A.~Garc\'ia}
\email{javier@caltech.edu}

\author[0000-0003-3828-2448]{Javier~A.~Garc\'ia}
\affil{Cahill Center for Astronomy and Astrophysics, California Institute of Technology, Pasadena, CA 91125, USA}
\affil{Dr. Karl Remeis-Observatory and Erlangen Centre for Astroparticle Physics, Sternwartstr.~7, 96049 Bamberg, Germany}

\author{John~A.~Tomsick}
\affil{Space Sciences Laboratory, 7 Gauss Way, University of California, Berkeley, CA 94720-7450, USA}

\author[0000-0002-5519-9550]{Navin~Sridhar}
\affil{Department of Astronomy, Columbia University, 550 W 120th St, New York, NY 10027, USA}

\author{Victoria~Grinberg}
\affil{Institut f\"ur Astronomie und Astrophysik (IAAT), Universit\"at T\"ubingen, Sand 1, D-72076 T\"ubingen, Germany}

\author{Riley~M.~T.~Connors}
\affil{Cahill Center for Astronomy and Astrophysics, California Institute of Technology, Pasadena, CA 91125, USA}

\author{Jingyi~Wang}
\affil{MIT Kavli Institute for Astrophysics and Space  Research, MIT, 70 Vassar Street, Cambridge, MA 02139}

\author{James~F.~Steiner}
\affil{MIT Kavli Institute for Astrophysics and Space  Research, MIT, 70 Vassar Street, Cambridge, MA 02139}

\author{Thomas~Dauser}
\affil{Dr. Karl Remeis-Observatory and Erlangen Centre for Astroparticle Physics, Sternwartstr.~7, 96049 Bamberg, Germany}

\author{Dominic~J.~Walton}
\affil{Institute of Astronomy, Madingley Road, Cambridge CB3 0HA, UK}

\author{Yanjun~Xu}
\affil{Cahill Center for Astronomy and Astrophysics, California Institute of Technology, Pasadena, CA 91125, USA}

\author{Fiona~A.~Harrison}
\affil{Cahill Center for Astronomy and Astrophysics, California Institute of Technology, Pasadena, CA 91125, USA}

\author{Karl~Foster}
\affil{Cahill Center for Astronomy and Astrophysics, California Institute of Technology, Pasadena, CA 91125, USA}

\author{Brian~Grefenstette}
\affil{Cahill Center for Astronomy and Astrophysics, California Institute of Technology, Pasadena, CA 91125, USA}

\author{Kristin~Madsen}
\affil{Cahill Center for Astronomy and Astrophysics, California Institute of Technology, Pasadena, CA 91125, USA}

\author{Andrew~Fabian}
\affil{Institute of Astronomy, Madingley Road, Cambridge CB3 0HA, UK}

%
%
%
%
%
%

\begin{abstract}
%
We report on the spectroscopic analysis of the black hole binary \gx339\
during its recent 2017--2018 outburst, observed simultaneously by the \swift\ and \nustar\
observatories. Although during this particular outburst the source failed to make state
transitions, and despite Sun constraints during the peak luminosity, we were able to trigger
four different observations sampling the evolution of the source in the hard state. We show
that even for the lowest luminosity observations the \nustar\ spectra show clear signatures of 
X-ray reprocessing (reflection) in an accretion disk. Detailed analysis of the highest signal-to-noise
spectra with our family of relativistic reflection models {\sc relxill} indicates the presence
of both broad and narrow reflection components. We find that a dual-lamppost model provides a superior
fit when compared to the standard single lamppost plus distant neutral reflection.
In the dual lamppost model two sources at different heights are placed on the rotational axis of
the black hole, suggesting that the narrow component of the Fe K emission is likely to 
originate in regions far away in the disk, but still significantly affected by its
rotational motions. Regardless of the geometry assumed, we find that the 
inner edge of the accretion disk reaches a few gravitational radii in all our fits, consistent with previous
determinations at similar luminosity levels. This confirms a very low degree of disk truncation
for this source at luminosities above $\sim 1$\% Eddington. Our estimates of \rin\ reinforces
the suggested behavior for an inner disk that approaches the inner-most regions as the luminosity
increases in the hard state.
\end{abstract}


%
%
%
%
\section{Introduction}\label{sec:intro}

The majority of stellar-mass black holes known to date are in low-mass X-ray binary
systems, which are transient in nature.
When in outburst, these black hole binary (BHB) systems
are readily observable in X-rays, as they display a rich phenomenology in the timing
and spectral domain. A typical BHB displays a fairly standard range of properties during a single
outburst, otherwise spending most of its time in a quiescent state \citep{rem06}. Outbursts can last 
months to years. During a mayor cycle a single BHB can show persistent and steady
jets, parsec-scale ballistic jets, quasi-periodic oscillations (QPOs) spanning 
0.01--450\,Hz, and transitions between spectral states broadly categorized as 
``hard" and ``soft", according to the overall slope of their X-ray continuum
\citep[e.g.,][]{fen04}. 

%
\begin{figure*}[ht!]
\centering
\includegraphics[width=\linewidth]{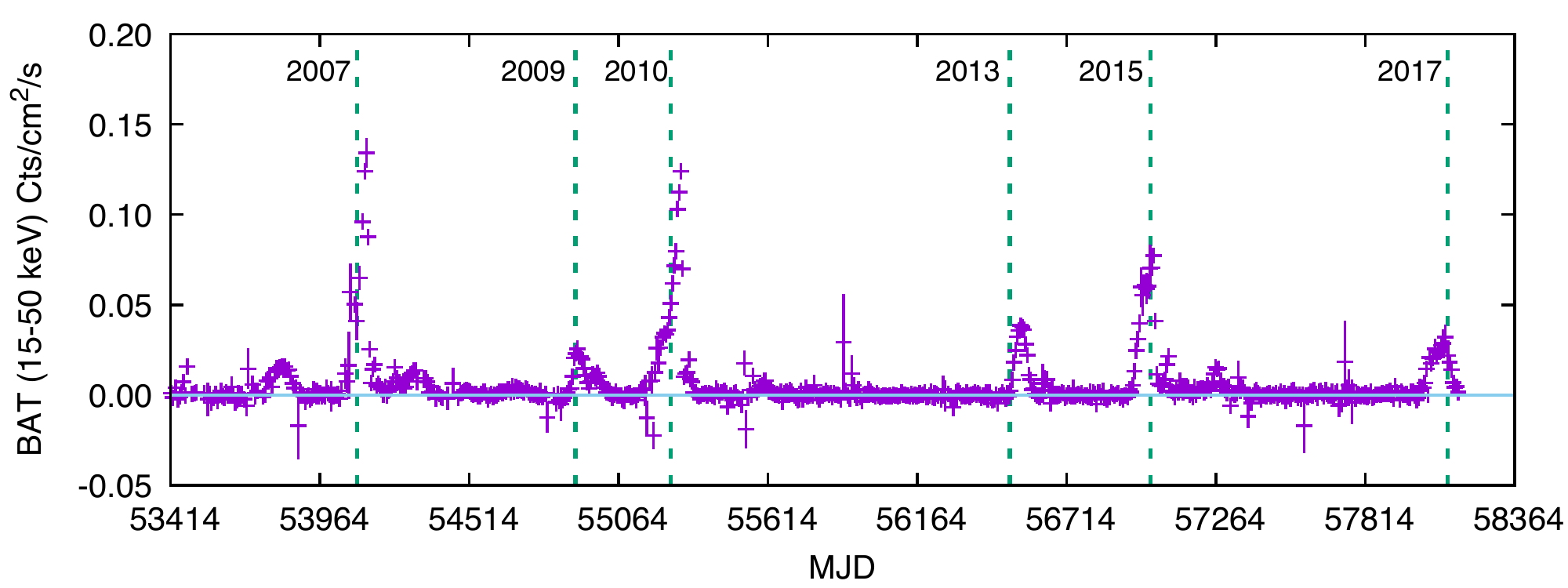}
\caption{
\swift\ BAT light curve (15--50\,keV) covering the last six outbursts
of \gx339, indicated with the vertical dashed lines. Similarly to 2013,
during the 2017 outburst the source failed to transition to the soft state.
}
\label{fig:full_LC}
\end{figure*}

In the hard state, the X-ray spectrum of BHBs is dominated by the non-thermal
emission produced by a hot ($T\sim10^8-10^9$\,K) and optically-thin ($\tau
\lesssim 1-2$) plasma referred to as the {\it corona}
\citep{haa93,dov97,zdz03}, in clear analogy to the Sun. The origin of the
corona is not well understood, but it has been associated with the base of a jet
\citep{mat92,mar05}. Recent global radiation magneto-hydrodynamic simulations
suggest that this corona can form naturally in suppermassive black holes
accreting at intermedium rates, i.e., 7\% to 20\% the Eddington rate
\citep{jiayf19}, although a definitive theoretical prediction has not been
fully produced.  This coronal radiation illuminates the accretion flow producing a
reflected spectrum containing many fluorescent atomic lines, absorption edges,
and other spectral features \citep{ros05,gar10}.  If the disk is close enough
to the black hole, the reflected spectrum is distorted by the strong
gravitational field \citep{fab89}. Thus, the precise modeling of these
signatures provides direct information on the state and composition of the
material in the disk, its geometry, inclination, location of the inner radius,
and ultimately the spin of the black hole \citep[e.g.,][]{dau10,rey14}.

\gx339\ is one of the most representative BHB systems known to date.
Classified as a low-mass X-ray binary, it shows full outbursts every 2--3 years
since its discovery \citep{mar73}. Even more frequent are the so called failed
outbursts, during which the source fails to transition to the soft state,
remaining in the hard state for the duration of a relatively shorter and less
luminous cycle.  In the last decade, \gx339\ has undergone three full outbursts
and about six failed ones.

We have previously analyzed archival data from the {\it Rossi X-ray Timing
Explorer} (\rxte), tracking the evolution of the hard state \citep{gar15}, and
the transition from hard to soft states (Sridhar et al., in prep.). However,
the proportional counter array (PCA) onboard of \rxte\ offers limited spectral
resolution. Meanwhile, several \nustar\ observations have covered the
hard-intermediate \citep{fue16} and soft-intermediate states \citep{par16}, and
the hard state at low luminosities during the failed outbursts of 2013
\citep{fue15} and 2015 \citep{wan18}. In this paper we report on the
observations of the 2017--2018 failed outburst of \gx339 \footnote{\gx339\
entered a new outburst on December 2018 \citep{gar18c}, which appears to
have also failed and is still in the decay phase as of the preparation of this
paper.}, as observed simultaneously with the \ngso\ \citep[\swift,][]{geh04},
and \nustar\ \citep{har13} instruments.

The remainder of the paper is organized as follows. In Section~\ref{sec:obs} we
provide the details of the observations analyzed here and the reduction
procedure. In Section~\ref{sec:anal} we present the main results of our
spectral analysis implementing relativistic reflection models. These results
are discussed in Section~\ref{sec:disc}, and our main conclusions are
summarized in Section~\ref{sec:concl}.

\section{Observations}\label{sec:obs}

The black hole binary \gx339\ is transient X-ray source with recurrent
outbursts every 2--3 years since its discovery in 1973.
Figure~\ref{fig:full_LC} shows the hard X-ray (15--50\,keV) lightcurve
covering the last six outbursts observed by the Burst Alert Telescope
\citep[BAT;][]{bar05} instrument onboard of 
\swift\ \citep{geh04}, provided by the Hard X-ray Transit Monitor \citep{kri13}.
Each outburst can reach a different peak luminosity, with the source
transitioning through all possible spectral accretion states \citep{tet16}.
However, in some cases the system remains in the hard state, which are often
referred as {\it failed outbursts} \citep[e.g.;][]{bux12,bel13,fue15}. These 
hard-only outbursts have been observed most recently in 2009, 2013, and lastly
in 2017.

%
\begin{deluxetable*}{ccccccc}[ht!]
\tablecaption{Observational Data Log for \gx339 \label{tab:obslog}}
\tablecolumns{7}
\tablewidth{0pt}
\tablehead{
\colhead{Epoch} & \colhead{Telescope}& \colhead{Instrument} & \colhead{ObsId} & \colhead{Date} & \colhead{Exp (ks)} & \colhead{Count Rate\tablenotemark{a}}
}
\startdata
1 & \nustar\  & FPMA/B & 80302304002 & 2017-10-02 & 23 & $2.1\pm 0.01$  \\
  & \swift\   & XRT    & 00032898149 & 2017-10-03 &  1 & $0.4\pm 0.02$  \\
2 & \nustar\  & FPMA/B & 80302304004 & 2017-10-25 & 21 & $22.8\pm 0.04$ \\
  & \swift\   & XRT    & 00032898155 & 2017-10-23 &  1 & $13.8\pm 0.14$ \\
3 & \nustar\  & FPMA/B & 80302304005 & 2017-11-02 & 23 & $16.9\pm 0.03$ \\
  & \swift\   & XRT    & 00032898160 & 2017-11-01 &  1 & $11.1\pm 0.13$ \\
4 & \nustar\  & FPMA/B & 80302304007 & 2018-01-30 & 32 & $3.2\pm 0.01$  \\
  & \swift\   & XRT    & 00032898163 & 2018-01-30 &  1 & $0.8\pm 0.03$  \\
\hline
\enddata
\tablenotetext{a}{Units are counts s$^{-1}$, measured in the 3--79\,keV band for \nustar,
and in the 0.5--10\,keV band for \swift. In the case of \nustar, the quoted value is for FPMA.}
\end{deluxetable*}
%

\begin{figure}[ht!]
\centering
\includegraphics[width=\linewidth]{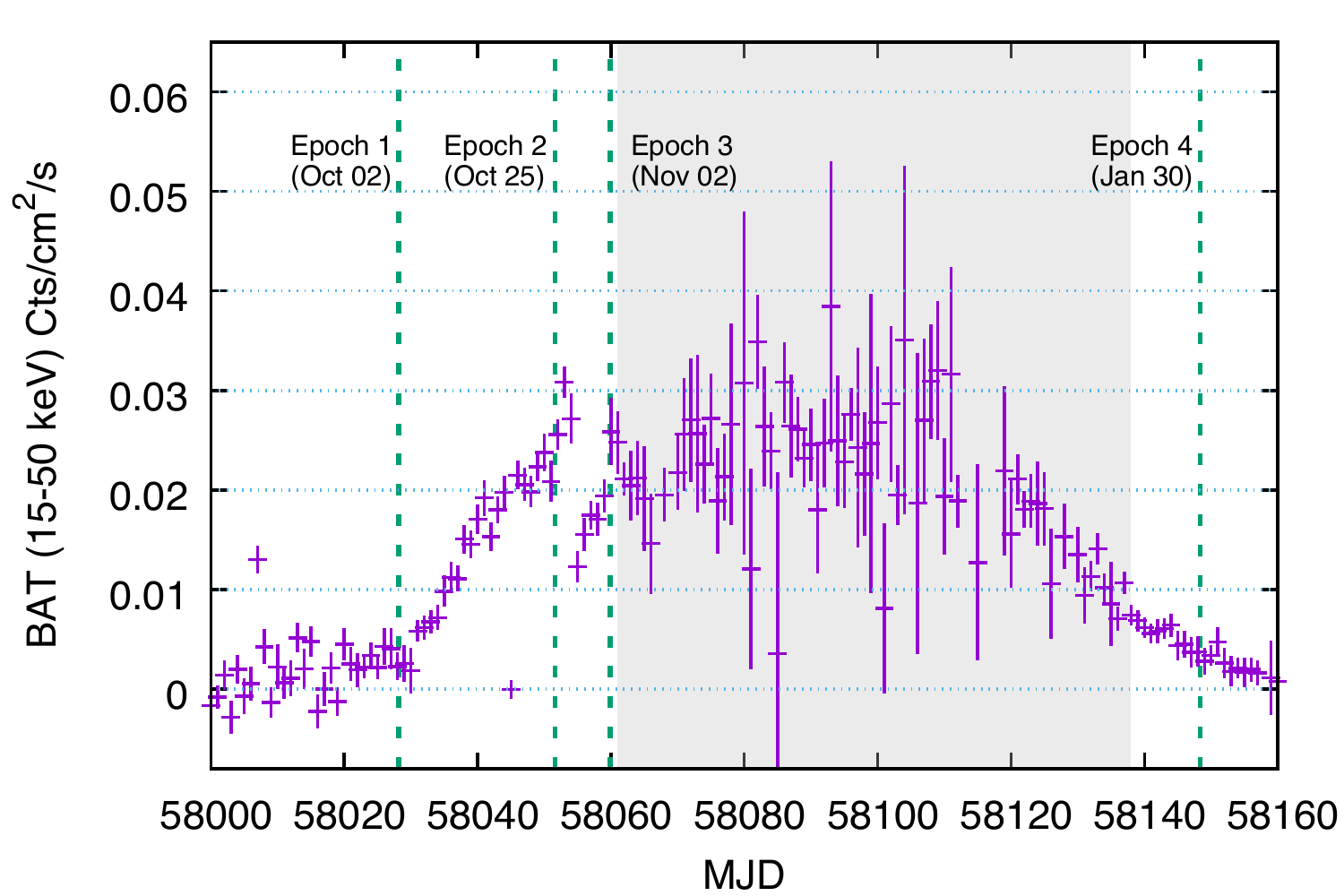}
\caption{
Closer look of the \swift\ BAT light curve (15--50\,keV) of \gx339\ during its
failed 2017 outburst. The vertical dashed lines indicate the starting times of
the \nustar\ observations (see Table~\ref{tab:obslog}). The shaded region indicates
the Sun constraint period during which observations were avoided.
}
\label{fig:LC}
\end{figure}

The last full outburst of \gx339\ was observed during 2014--2015. A new
outburst was reported after a brightening observed in the optical band on 2017,
September \citep{rusd17}, followed by detections in X-rays with \swift\
\citep{gan17}, as well as in the radio with the Australia Telescope Compact
Array \citep[ATCA;][]{rust17}. Following these detections, we triggered a
series of Target of Opportunity observations from our Guest Observer program
with the {\it Nuclear Spectroscopic Telescope Array} \citep[\nustar;][]{har13}.
Figure~\ref{fig:LC} shows the \swift\ BAT lightcurve around the time of the
2017 outburst, during which four \nustar\ and \swift\ XRT observations were
performed. The first Epoch was triggered on 2017 October 2 03:40:06 UT
detecting the source at $\sim 3.3$\,mCrab\footnote{1\,mCrab $=1.4\times
10^{-11}$\,erg\,cm$^{-2}$\,s$^{-1}$ \citep[2--10\,keV;][]{kir05}. Flux based on
a simple absorbed power-law model.} \citep[2--10\,keV;][]{gar17}. Two other
observations were triggered during the rise of the outburst, until the source
entered a Sun constrained period for \nustar\ (shown with the shaded region in
Figure~\ref{fig:LC}).  The last observation took place on 2018 January 30, at
which point the X-ray flux had already decreased to a level similar to that
observed during Epoch~1 ($\sim 5$\,mCrab, 2--10\,keV). Details of all the
Epochs analyzed in this paper are summarized in Table~\ref{tab:obslog}.

\subsection{NuSTAR Extraction}

The four \nustar\ observations listed in Table~\ref{tab:obslog} were reduced
using the standard pipeline Data Analysis Software ({\sc nustardas}, v1.8.0),
in combination with the {\sc caldb} instrumental calibration files v20170817, which
are part of {\sc heasoft} v6.24. Event files were cleaned using standard filtering
parameters using the {\sc nupipeline} task, reducing internal background and 
removing data taken close to the South Atlantic Anomaly. Source products (spectra
and lightcurves), backgrounds, and instrumental responses were produced for
each Focal Plane Module A and B (FPMA/FPMB) using the {\sc nuproducts} task.
The source products were extracted from a 60" circular region centered at the
position of \gx339, while background spectra were extracted from a 100" region
placed at the opposite side of the same detector. Finally, spectra were binned
requiring a minimum signal-to-noise of 5 per channel, and to oversample the 
instrument's resolution by a factor of 3. We fitted the FPMA/B spectra in the
entire energy range (3--79\,keV).

%

\subsection{Swift Extraction}

We made energy spectra for the four \swift\ XRT observations listed in
Table~\ref{tab:obslog}  using HEASOFT v6.21 and version x20150721 of the XRT calibration files.
For the observations associated with Epochs~1 and 4, XRT was in photon counting
(PC) mode, and XRT was in windowed timing (WT) mode for Epochs~2 and 3.
Although the count rates were lower for the PC mode observations, they were
still high enough for the region at the center of the point spread function to
be subject to photon pile-up, and we extracted these two spectra from an
annulus with inner radius of $10^{\prime\prime}$ and an outer radius of
$47^{\prime\prime}$.  For the WT mode observations, the extraction region was
simply a circle with a radius of $47^{\prime\prime}$.  In all four cases, we
subtracted background by making a spectrum from counts in an annulus around the
source.  We used the response matrix file swxpc0to12s6\_20130101v014.rmf from
the XRT calibration database and {\tt xrtmkarf} with exposure maps
created for each observation to produce the ancillary response file.  Finally,
the spectra were binned to produce the final 0.5-10\,keV spectra that we used
for our analysis.

\section{Spectral Analysis}\label{sec:anal}

We analyze the time-averaged spectra from the Epochs taken with \nustar\ and
\swift\ observatories. Although these observations are not strictly
simultaneous, they were taken within 1--2 days of each other.  Epochs 1 and 4
were taken at the beginning and end of the failed outburst, coincidentally at
similar flux levels. Meanwhile, Epochs 2 and 3 were taken close to the peak of
the outburst right before the observations were constrained by the source being
too close the Sun, also with similar fluxes (see Figure~\ref{fig:LC}). All the 
fits and statistical analysis is performed with the spectral package {\sc xspec}
\cite[v12.10.0c;][]{arn96}. 

%
\begin{figure*}[ht!]
\centering
\includegraphics[width=0.49\textwidth,trim={0 1.5cm 0 0}]{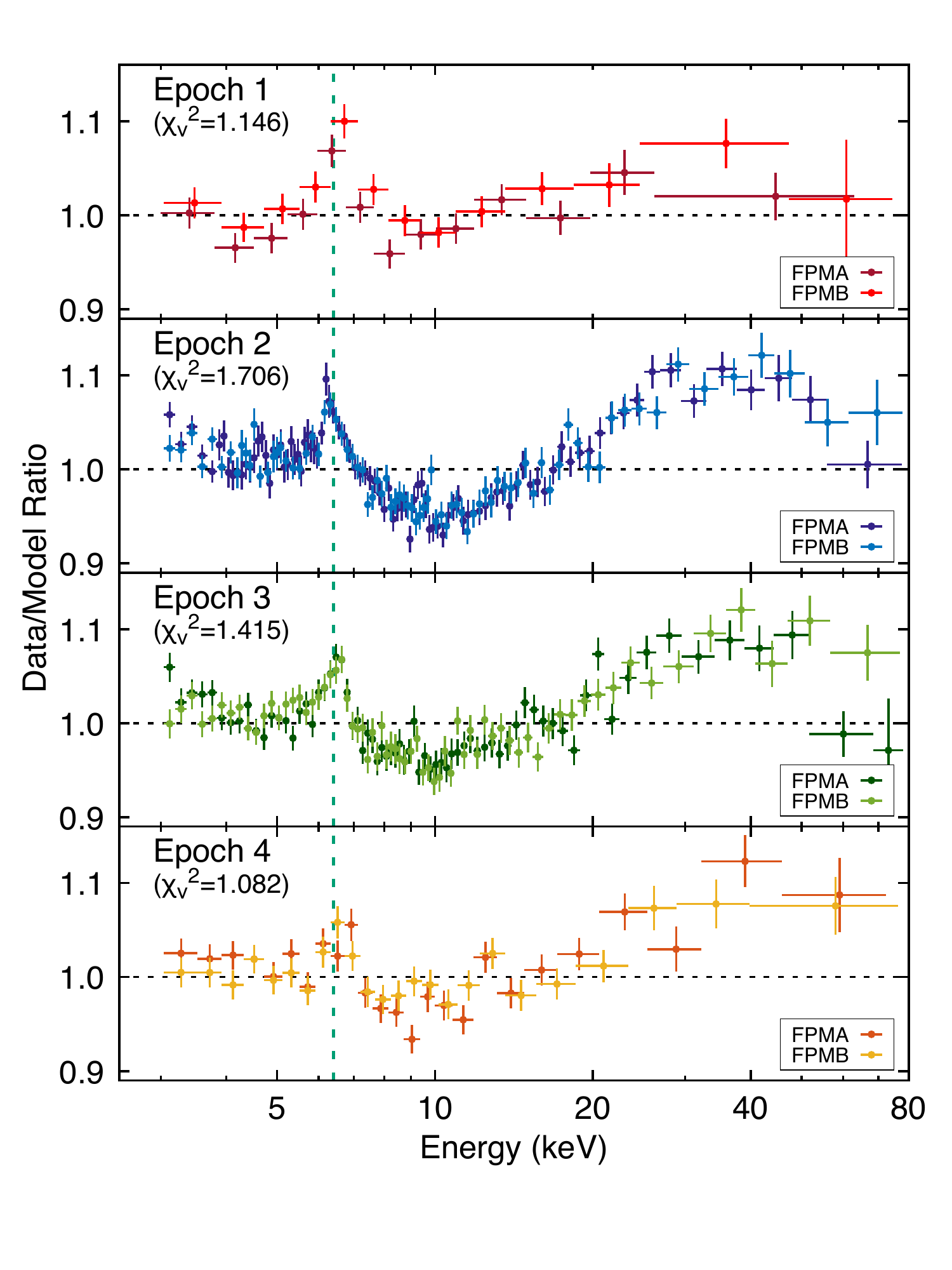}
\includegraphics[width=0.49\textwidth,trim={0 1.5cm 0 0}]{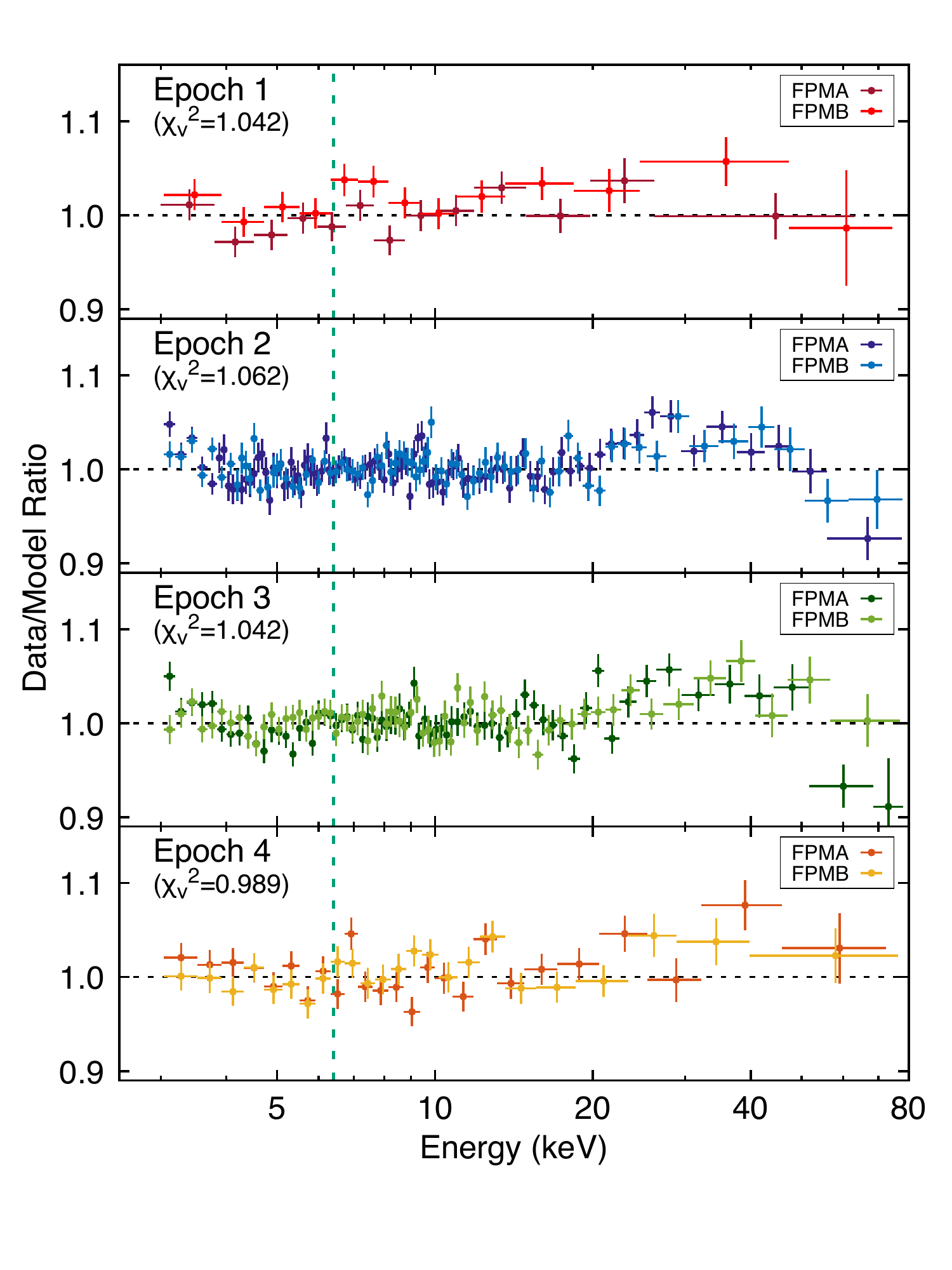}
\caption{
({\it left:}) Data-to-model ratio plots of the \nustar\ observations fitted to
an absorbed Comptonization continuum model ({\tt TBabs*nthComp}). The four
spectra show clear signatures of reflection, including Fe K emission near
6.4\,keV (vertical dashed line), K-edge, and Compton hump.  Spectra have been
further rebinned to improve clarity. ({\it right:}) Ratio plots of fits
including a Gaussian emission profile and a smeared edge component ({\tt
TBabs*nthComp*smedge*gau}).  For Epochs~2 and 3, this model cannot fully
reproduce the Compton hump above $\sim 20$\,keV.  However, the marked
improvement in the fit-statistics indicates the significance of the reflection
features.
}
\label{fig:ratios}
\end{figure*}

\subsection{Simple Description of the NuSTAR Data}

We first fit a simple absorbed continuum model to all the \nustar\ data, using
the thermal Comptonization model {\tt nthComp} \citep{zdz96,zyc99}.
In {\sc xspec} notation this model is written as
{\tt TBabs*nthComp}, where {\tt TBabs} accounts for the neutral photoelectric
absorption in the intergalactic medium, using the cross sections by
\cite{ver96} and the cosmic abundances by \cite{wil00}. For simplicity, at this
point we fixed the hydrogen column density to a value similar to that found in
previous studies \citep[$N_\mathrm{H}=6\times10^{21}$\,cm$^{-2}$,][]{gar15,wan18}.
This parameter will be investigated later and allowed to vary freely in the final
fits.  The {\tt nthComp} component
describes the power-law like continuum produced by Comptonization of thermal
disk photons in a hot gas of electrons \citep{zdz96,zyc99}. 

Figure~\ref{fig:ratios} (left panels)
shows ratio plots of these fits. Clear signatures of reflection are observed
for all the spectra: Fe K-shell emission near 6--7\,keV,
Fe K-edge absorption near 10\,keV, and a Compton hump peaking near 40\,keV.
Epochs 2 and 3 have the highest signal-to-noise and thus the most significant
detection of the reflection spectrum. However, reflection is also evident in
Epochs 1 and 4 despite the flux being lower by roughly an order of magnitude.
The resulting
reduced $\chi^2$ (shown in each panel) can be interpreted as an indication of the
significance for the detection of the reflection signal.

A functional characterization of the reflection signatures can be achieved by
fitting a Gaussian profile for the Fe K emission, and smeared edge component
\citep[{\tt smedge};][]{ebi91} for the Fe K-edge. The {\tt Gaussian} component is fitted
with a fixed width of $\sigma=0.01$\,keV, while the energy and normalization are
free to vary. The {\tt smedge} component was fitted with fixed index (-2.67) and
width (7), while the energy and optical depth were let free to vary.
  The best-fit parameters are
summarized in Table~\ref{tab:fitsI}, together with an estimation of the
equivalent width (EW) for the Gaussian profile, while ratio plots are shown in 
the right panels of Figure~\ref{fig:ratios}. It is evident that the inclusion of these
components significantly improves the fit in all the four Epochs, with a clear
decrease of the reduced chi-squared $\chi_{\nu}^2 = \chi^2/\nu$ (comparing with the
values quoted in the left panels of Figure~\ref{fig:ratios}),
where $\nu$ is the number of degrees of freedom (d.o.f). Inspection of the residuals
also shows that this simple functional model cannot fully reproduce the shape of
the Compton hump above $\sim 20$\,keV, at least the case of Epochs~2 and 3. However,
given the limited signal of the data at those energies, the penalty on the fit
statistics is minimal.

\subsection{Analysis of the NuSTAR and Swift Data: Epochs~1 and 4}

We now turn to a physically motivated model to describe the spectra of \gx339\ during its
2017 failed outburst. As shown above, there are clear signatures of X-ray reprocessing in an
optically-thick atmosphere. Thus, we use the reflection model \relxill\ \citep{gar14a,dau14} to fit these data.
However, given their limited signal, data from Epochs~1 and 4 do not provide good constraints.
In these cases, a simple reflection without any relativistic effects included \citep[modeled with \xillver,][]{gar10,gar13a}
already provides a good fit of the spectra. Testing a relativistic version of the same model
(e.g., \relxill) provides an equivalent fit, but with most parameters poorly constrained.
This is likely due to an over-description of the data, given the many free parameters of the model.
At most, when fitted with relativistic reflection, these data suggest that the accretion disk
may be significantly truncated, as no broad component of the Fe K line can be detected.
Nevertheless, it is unclear if the lack of a broad component is real, or simply due to the
poor statistics of the data. Thus, we are cautious in drawing any definitive conclusions from
these two spectra. Therefore, for the remainder of the analysis, we concentrate on Epochs~2
and 3, the two spectra taken during the brightest part of this outburst.

\def\fIgamEI{$1.605 \pm 0.009$}		
\def\fInorEI{$0.17 \pm 0.01$}	
\def\fIeneEI{$6.41 \pm 0.05$}		
\def\fInogEI{$0.6 \pm 0.1$}		
\def\fIensEI{$7.2 \pm 0.7$}		
\def\fItauEI{$0.2 \pm 0.1$}		
\def\fIredchiEI{$1.042$}		
\def\fIeqwEI{$74 \pm 26$}		
\def\fIgamEII{$1.565 \pm 0.004$}        
\def\fInorEII{$1.4 \pm 0.01$}         
\def\fIeneEII{$6.34 \pm 0.05$}          
\def\fInogEII{$1.9 \pm 0.4$}          
\def\fIensEII{$6.9 \pm 0.1$}          
\def\fItauEII{$0.63 \pm 0.04$}          
\def\fIredchiEII{$1.062$}		
\def\fIeqwEII{$23 \pm 8$}		
\def\fIgamEIII{$1.550 \pm 0.005$}       
\def\fInorEIII{$1.2 \pm 0.01$}        
\def\fIeneEIII{$6.53 \pm 0.05$}         
\def\fInogEIII{$1.9 \pm 0.4$}         
\def\fIensEIII{$6.7 \pm 0.2$}         
\def\fItauEIII{$0.49 \pm 0.05$}         
\def\fIredchiEIII{$1.042$}		
\def\fIeqwEIII{$28 \pm 10$}		
\def\fIgamEIV{$1.602 \pm 0.006$}        
\def\fInorEIV{$0.25 \pm 0.01$}       
\def\fIeneEIV{$6.4 \pm 0.1$}           
\def\fInogEIV{$0.3 \pm 0.2$}           
\def\fIensEIV{$6.9 \pm 0.3$}           
\def\fItauEIV{$0.40 \pm 0.08$}          
\def\fIredchiEVI{$0.989$}		
\def\fIeqwEIV{$ < 36$}			

%
\begin{table*}
\begin{center}
\small
\caption{Best-fit parameters for the final fits with relativistic reflection modeling.}
\begin{tabular}{lcrrrr}
\hline
Component & Parameter & Epoch~1 & Epoch~2 & Epoch~3 & Epoch~4 \\
\hline
{\tt TBabs}    & $N_\mathrm{H}$ ($10^{21}$ cm$^{-2}$) &  6  &  6  &  6  &  6  \\
{\tt nthComp}  & $\Gamma$                             & \fIgamEI & \fIgamEII & \fIgamEIII & \fIgamEIV \\
{\tt nthComp}  & $N$\tablenotemark{a} ($10^{-1}$)                      & \fInorEI & \fInorEII & \fInorEIII & \fInorEIV \\
{\tt Gaussian} & $E$ (keV)                            & \fIeneEI & \fIeneEII & \fIeneEIII & \fIeneEIV \\
{\tt Gaussian} & $N$ ($10^{-4}$\,photons\,cm$^{-2}$\,s$^{-1}$)                      & \fInogEI & \fInogEII & \fInogEIII & \fInogEIV \\
{\tt Smedge}   & $E$ (keV)                            & \fIensEI & \fIensEII & \fIensEIII & \fIensEIV \\
{\tt Smedge}   & Max $\tau$                           & \fItauEI & \fItauEII & \fItauEIII & \fItauEIV \\
\hline
$\chi_{\nu}^2$ & \nodata                              & \fIredchiEI &\fIredchiEII &\fIredchiEIII &\fIredchiEVI \\
EW (eV)        & \nodata                              & \fIeqwEI &\fIeqwEII &\fIeqwEIII &\fIeqwEIV \\
\hline
\end{tabular}
\tablenotetext{a}{When normalization is equal to unity, the model flux is 
1\,photon\,keV$^{-1}$\,cm$^{-2}$\,s$^{-1}$ at 1\,keV.}
\label{tab:fitsI}
\end{center}
\end{table*}

\subsection{Analysis of the NuSTAR and Swift Data: Epochs~2 and 3}

\begin{figure*}[ht!]
\centering
\includegraphics[width=0.45\linewidth,trim={0 1cm 0 0}]{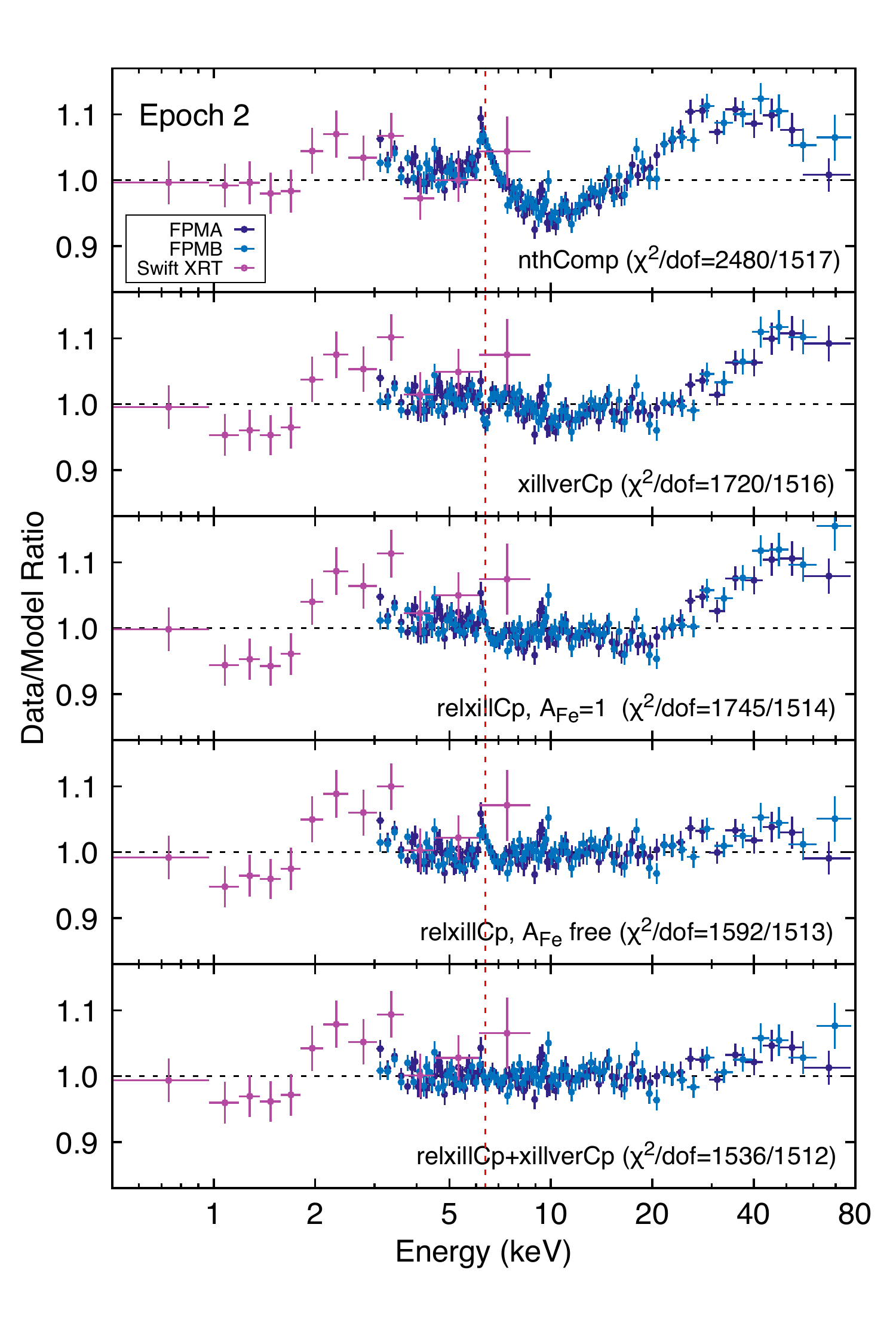}
\includegraphics[width=0.45\linewidth,trim={0 1cm 0 0}]{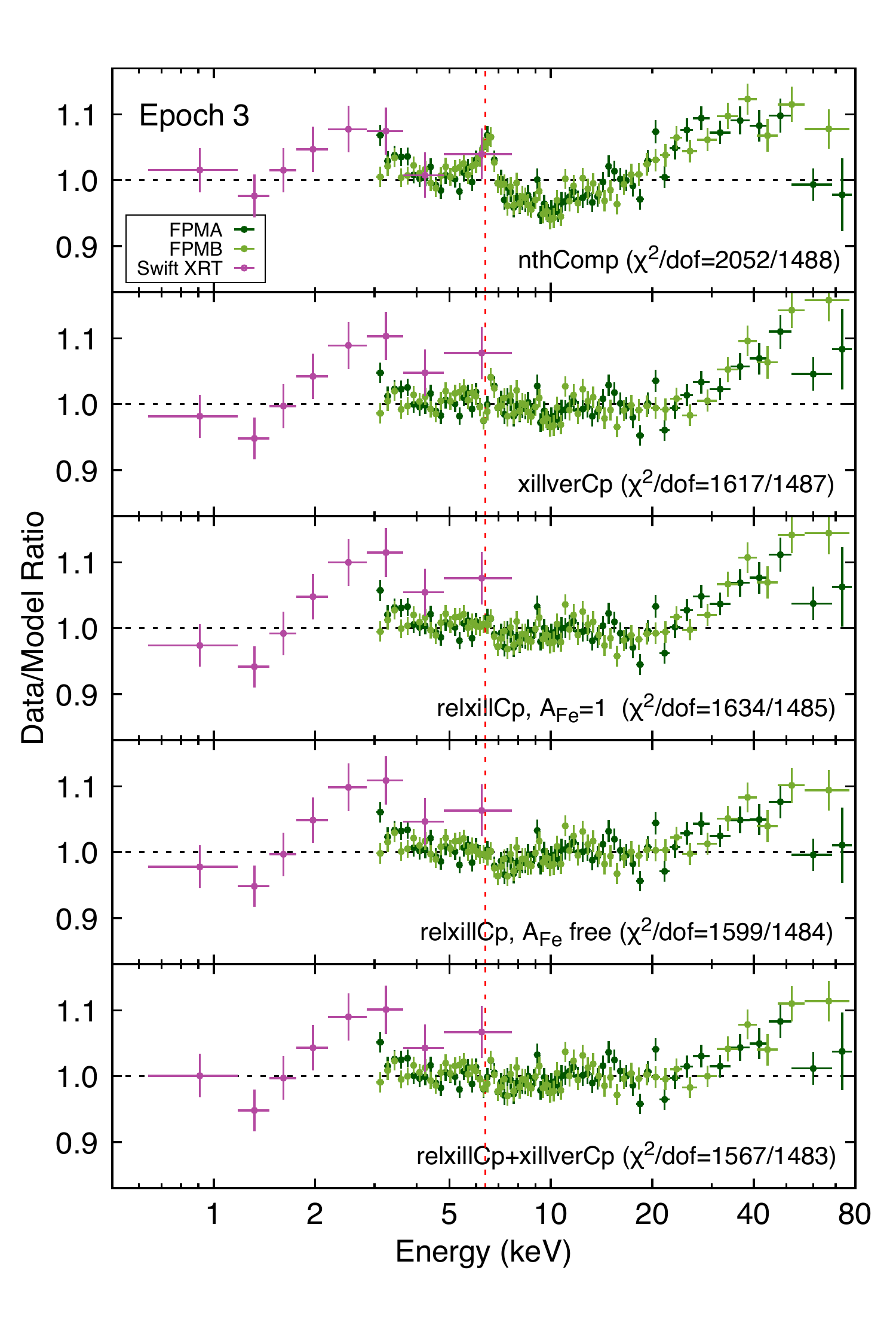}
\caption{
Data-to-model ratio plots of the \swift\ and \nustar\ spectra fitted with a progression of
different models of increasing complexity. The models are indicated on each panel, starting with
a simple absorbed Comptonization continuum model ({\tt TBabs*nthComp}), and then different flavors
of our reflection model \relxill. Also indicated are the fit statistics for each case.
Left and right panels correspond to Epochs~2 and 3, respectively.
The vertical dashed line indicates 6.4\,keV.
}
\label{fig:ratios23}
\end{figure*}

The simple fits to the \nustar\ data described above revealed the presence of
clear reflection signatures. Furthermore,
the inclusion of \swift\ XRT data shows a relatively weak excess flux in the
soft bands, which can be modeled with a thermal-disk emission model \citep[{\tt diskbb};][]{mit84}.
We have investigated these components further by
following a progression of different model combinations, fitting the \swift\
and \nustar\ data simultaneously for Epochs~2 and 3.  Ratio plots resulting from this
progression of different reflection model components are shown in
Figure~\ref{fig:ratios23}. We start with a model to describe the disk emission
plus the non-thermal power-law continuum with {\tt TBabs*(diskbb+nthComp)}.
The hydrogen column is held fixed to $N_\mathrm{H}=6\times10^{21}$\,cm$^{-2}$,
as at this point we are only interested in an overall comparison of the
reflection components. In all our fits the electron temperature is pegged at
its maximum allowed value of $kT_e = 400$\,keV. This parameter, which represents
the temperature of the electrons in the Comptonizing corona, is constrained
mostly by the cutoff of the power-law continuum at high energies, which is
roughly at $E_\mathrm{cut}\sim 2-3 (kT_e)$.  The fact that we cannot detect
such a curvature means that the cutoff must at the very least at $\sim
800$\,keV or above. This is consistent with the results in \cite{gar15}, where
we found $E_\mathrm{cut} \gtrsim 800$\,keV for observations at $\sim 2-4$\% of
the Eddington luminosity. Thus, in the fits presented here, we fix the electron
temperature at its upper limit of 400\,keV in the {\tt nthComp} component,
which is also linked to the same parameter in the reflected component.
Meanwhile, the disk seed-photon temperature is linked to that in the {\tt diskbb}
component.  As in the case of the \nustar\ data alone, this model provides a
decent fit to the continuum (\redcsq=1.64 and 1.38; for Epochs~2 and 3,
respectively), but with clear residuals near 6.4\,keV (reminiscent of Fe K
emission) and near 30--40\,keV (reminiscent of the Compton hump).

Thus, we first attempted to describe the observed residuals with a single reflection
component with the \xillverCp\ model \citep{gar10,gar13a}. This model produces an 
ionized reflection spectrum produced by the illumination of an optically-thick
slab with a fixed gas density of $n_e = 10^{15}$\,cm$^{-3}$. The hard X-ray continuum
is assumed to be produced by thermal Comptonization of disk photons in a hot corona,
and the spectrum is calculated using the {\tt nthComp} model \citep{zdz96,zyc99}.
The photon index and electron temperature are linked to those in {\tt nthComp}. We
further assumed a nearly neutral gas ($\log\xi=0$) and solar iron abundance ($A_\mathrm{Fe}=1$).
Finally, the inclination is also fixed to $40$\,deg, a value typically found with
reflection modeling of \gx339\ \citep[][see discussion at the end of this Section]{gar15,wan18}.
The improvement of the fit is very dramatic, with $\Delta\chi^2=760$ (Epoch~2) and $435$ 
(Epoch~3), for only one extra d.o.f. in both cases. This clearly
shows the high significance of the reflection features. However, strong residuals
are still observed at high energies ($>20$\,keV), and the Fe K emission appears to be
over-estimated (Figure~\ref{fig:ratios23}). Allowing the ionization parameter in 
the \xillverCp\ component to vary
freely does not provide a statistically significant improvement of the fit ($\chi^2$ 
only reduces by $\sim 2$). As the narrow component of the Fe K emission peaks near
$6.4$\,keV, this is likely driving the model toward a neutral-like reflection.

%

We then replaced the reflection component with its relativistic counterpart \relxillCp\ 
\citep{gar14a,dau14}. In this model, the reflection spectrum is convolved with a general
relativistic kernel to account for the distortion effects caused by the strong gravitational
field near the black hole. In this variant, the emissivity of the disk is assumed to
follow a power-law with radius ($\propto r^q$). For simplicity, we kept the emissivity
index fixed to the canonical value $q=3$, which describes the profile in a standard 
\cite{sha73} disk. The spin of the black hole is assumed to be
at maximum ($a_* = 0.998$), and again iron abundance is assumed to be solar. Only the
location of the disk inner radius \rin\ was allowed to vary. This model provides a slightly
worse fit than the one using \xillverCp\, and it cannot reproduce the narrow component
of the Fe K emission despite pushing the inner radius to $\sim 100$\,\risco. Most
importantly, the strongest residuals at high energies remained unmodified.

In the next progression we then tried the same relativistic reflection model but allowing
the iron abundance to be free. This parameter alone significantly improves the fit,
with $\Delta\chi^2=153$ and $35$ for Epochs~2 and 3, respectively. This is mostly due to 
a much better fit of the high-energy part of the spectrum.  The required abundance
is significantly larger than the solar value ($A_\mathrm{Fe}\sim 5$), which is consistent
with previous studies of this source \citep[e.g.,][]{fue15,gar15,wan18}.
This suggests that the specific shape of the Compton
hump around 10--40\,keV is mostly determined by the amount of iron in the gas.
We also notice that the inner radius is significantly smaller than in the fit with
fixed solar abundance (\rin$\sim 2-5$\,\risco). However, as before the narrow component
of the Fe K emission is still not well modeled.

Therefore, the final progression is to include a distant reflector (i.e., not affected
by relativistic effects) together with the relativistic one. This is done by using both 
of the \relxillCp\
and \xillverCp\ models, which provides the best fit of the data in this progression
of models (\redcsq=1.016 for Epoch~2; and \redcsq=1.060 for Epoch~3). All the major 
features are well described by this model, including the broad and narrow components
of the Fe K emission, the Fe K-edge, and the Compton hump. 

We note that our particular choice for the inclination has a non-negligible effect in the 
overall fits, but it is not arbitrary. When performing the fits with free inclination,
we found this parameter to be loosely constrained, but with preference for very low values
(close it its lower limit of 3 deg). Low inclinations ($i \lesssim 30$\,deg) can be easily
ruled out by considering the most recent measurements of the mass function for this system,
provided by \cite{hei17}. Meanwhile, most recent reflection spectroscopy analyses are
broadly consistent with the inclination reported for this source \citep[e.g.,][]{gar15,par16,bas16},
which are also well within the range derived by \cite{zdz19} on theoretical arguments.
In a more extensive work, \cite{dzi19} presented a systematic analysis of a high S/N
spectrum for \gx339\ in the hard state \citep[specifically, the same as in Box~B in][]{gar15},
using several different combinations of reflection models, arguing that results can be
model dependent. However, the inclination found in their fits is either consistent with 
$\sim40$\,deg, or poorly constrained (see their Table~1). The largest discrepancies in the 
inclination derived from reflection spectroscopy is found with respect to earlier works
by \cite{mil06} ($i=20^{+5}_{-15}$\,deg), and \cite{rei08} ($i\lesssim 20$\,deg). Both 
of these works fitted \xmm\ data only (thus, no high-energy coverage), using outdated
reflection models \cite[see discussion in Section~6.1.2 of][]{gar15}.
We have chosen the value of 40\,deg, which was obtained in our most recent analysis of this
source using the same instruments (\nustar\ and \swift), with observations taken at a relatively
similar accretion state, and fitted with the same reflection models \citep{wan18}.

\subsection{Relativistic and Distant Reflection vs. Dual-Lamppost Models}

This simple progression of 
models discussed above indicates that both relativistic (inner regions) and non-relativistic (distant)
reflection components are needed to fit these data, and that the complexity of the
Fe K-edge and Compton hump region requires an iron abundance significantly larger than
solar. Based on these findings, we now explore some additional and more detailed scenarios in which
a lamppost geometry is assumed for the illumination
of the disk (\relxilllpCp). In this setup, the primary photons originate from a point
source located on the spin axis at some height $h$ above the black hole. As the
geometry is prescribed, the model can self-consistently calculate the reflection fraction
(and reflection strength) given a set of spin, inner radius, and coronal height. Note that
the reflection fraction is independent of the inclination of the reflector, as it is 
defined as the ratio of the coronal intensity that reaches the disk to the coronal intensity
that reaches an observer at infinity directly \citep{dau16}. We refer to this
as Model~A, which in {\sc xspec} language is written as
\\

\def\fIIEIIa{$0.998$}			 	 
\def\fIIEIIinc{$40$}			 	 
\def\fIIEIIrout{$1000$}			 	 
\def\fIIEIIkte{$400$}			 	 
\def\fIIEIIdg{$-0.04$}		         
\def\fIIEIInh{$6.2^{+0.8}_{-0.5}$}		 
\def\fIIEIIh{$1.6^{+0.1}_{-0.2}$}		 
\def\fIIEIIrin{$< 1.2$}			 
\def\fIIEIIga{$1.37^{+0.02}_{-0.01}$}	 
\def\fIIEIIlxi{$4.06^{+0.01}_{-0.01}$}		 
\def\fIIEIIafe{$6.5^{+0.7}_{-2.7}$}		 
\def\fIIEIIrxn{$8^{+5}_{-4}$}		 
\def\fIIEIIxin{$1.5^{+0.1}_{-0.2}$}		 
\def\fIIEIItin{$89^{+24}_{-10}$}		 
\def\fIIEIIdin{$< 2$}		 
\def\fIIEIIdga{$2.4^{+0.6}_{-0.6}$}		 
\def\fIIEIIcrni{$0.97^{+0.01}_{-0.01}$}		 
\def\fIIEIIcrnii{$0.94^{+0.02}_{-0.02}$}	 
\def\fIIEIIrf{$8.54$}			 	 
\def\fIIEIIrs{$2.13$}			 	 
\def\fIIEIIpf{$45.8$}			 	 
\def\fIIEIIchi{$1531.8$}			 
\def\fIIEIIdof{$1511$}				 
\def\fIIEIIred{$1.014$}				 

\def\fIIEIIInh{$6.3^{+0.4}_{-0.4}$}		 
\def\fIIEIIIh{$1.6^{+0.04}_{-0.2}$}		 
\def\fIIEIIIrin{$< 1.2$}			 
\def\fIIEIIIga{$1.37^{+0.02}_{-0.02}$}	 
\def\fIIEIIIlxi{$4.07^{+0.02}_{-0.01}$}		 
\def\fIIEIIIafe{$6.7^{+0.7}_{-1}$}		 
\def\fIIEIIIrxn{$6^{+12}_{-2}$}		 
\def\fIIEIIIxin{$1.1^{+0.1}_{-0.2}$}		 
\def\fIIEIIItin{\nodata}			 
\def\fIIEIIIdin{\nodata}			 
\def\fIIEIIIdga{$0.2^{+0.3}_{-0.7}$}		 
\def\fIIEIIIcrni{$1.03^{+0.01}_{-0.02}$}	 
\def\fIIEIIIcrnii{$0.87^{+0.02}_{-0.02}$}	 
\def\fIIEIIIrf{$8.30$}                            
\def\fIIEIIIrs{$2.23$}                            
\def\fIIEIIIpf{$42.8$}                            
\def\fIIEIIIchi{$1569.8$}			 
\def\fIIEIIIdof{$1486$}				 
\def\fIIEIIIred{$1.056$}			 

%
\begin{table*}
\begin{center}
\small
\caption{Best-fit parameters for Epochs~2 and 3 using a lamppost plus a distant reflection component
(Model~A; {\tt crabcorr*TBabs*(diskbb+relxilllpCp+xillverCp}). }
\begin{tabular}{lcrr}
\hline
Component & Parameter & Epoch~2 & Epoch~3 \\
\hline
{\tt relxilllpCp} & $a_*$                                & \multicolumn{2}{c}{\fIIEIIa\ }    \\
{\tt relxilllpCp} & $i$ (deg)                            & \multicolumn{2}{c}{\fIIEIIinc\ }  \\
{\tt relxilllpCp} & $R_\mathrm{out}$ ($R_\mathrm{g})$    & \multicolumn{2}{c}{\fIIEIIrout\ } \\
{\tt relxilllpCp} & $kT_\mathrm{e}$ (keV)                & \multicolumn{2}{c}{\fIIEIIkte\ }  \\
{\tt crabcorr}    & $\Delta\Gamma$ (XRT)                 & \multicolumn{2}{c}{\fIIEIIdg\ }   \\
\hline
{\tt TBabs}       & $N_\mathrm{H}$ ($10^{21}$ cm$^{-2}$) & \fIIEIInh\  & \fIIEIIInh\  \\
{\tt relxilllpCp} & $h$ $(R_\mathrm{Hor})$               & \fIIEIIh\   & \fIIEIIIh\   \\
{\tt relxilllpCp} & $R_\mathrm{in}$ ($R_\mathrm{ISCO})$  & \fIIEIIrin\ & \fIIEIIIrin\ \\
{\tt relxilllpCp} & $\Gamma$                             & \fIIEIIga\  & \fIIEIIIga\  \\
{\tt relxilllpCp} & $\log\xi$ (erg cm s$^{-1}$)          & \fIIEIIlxi\ & \fIIEIIIlxi\ \\
{\tt relxilllpCp} & $A_\mathrm{Fe}$                      & \fIIEIIafe\ & \fIIEIIIafe\ \\
{\tt relxilllpCp} & $N$ $(10^{-2})$\tablenotemark{a}     & \fIIEIIrxn\ & \fIIEIIIrxn\ \\
{\tt xillverCp}   & $N$ $(10^{-3})$\tablenotemark{a}     & \fIIEIIxin\ & \fIIEIIIxin\ \\
{\tt diskbb}      & $kT_\mathrm{in}$ (eV)                & \fIIEIItin\ & \fIIEIIItin\ \\
{\tt diskbb}      & $N$ $(10^{7})$\tablenotemark{b}      & \fIIEIIdin\ & \fIIEIIIdin\ \\
{\tt crabcorr}    & $\Delta\Gamma$ ($10^{-2}$, FPMB)     & \fIIEIIdga\   & \fIIEIIIdga\   \\
{\tt crabcorr}    & $N$ (FPMB)                           & \fIIEIIcrni\  & \fIIEIIIcrni\  \\
{\tt crabcorr}    & $N$ (XRT)                            & \fIIEIIcrnii\ & \fIIEIIIcrnii\ \\
\hline
Reflection Fraction & $R_\mathrm{f}$                     & \fIIEIIrf\  & \fIIEIIIrf\  \\
Reflection Strength & $R_\mathrm{s}$                     & \fIIEIIrs\  & \fIIEIIIrs\  \\
Photons Lost in BH  & $N_\mathrm{ph}$ (\%)               & \fIIEIIpf\  & \fIIEIIIpf\  \\
\hline
\multicolumn{2}{c}{$\chi^2$}                             & \fIIEIIchi\ & \fIIEIIIchi\ \\ 
\multicolumn{2}{c}{$\nu$}                                & \fIIEIIdof\ & \fIIEIIIdof\ \\
\multicolumn{2}{c}{$\chi_{\nu}^2$}                       & \fIIEIIred\ & \fIIEIIIred\ \\
\hline
\end{tabular}
\tablenotetext{a}{The normalization of \xillver\ is defined such that for a source spectrum
with flux $F_\mathrm{x}(E)$ incident on a disk with density $n_e$, then 
$\int_{0.1\mathrm{keV}}^{1\mathrm{MeV}}{F_\mathrm{x}(E)dE}=10^{20}\frac{n_e\xi}{4\pi}$, where
$\xi$ is the ionization parameter. For the \relxill\ models the definition is identical, although 
the observed flux differs due to the relativistic effects \citep[see][]{dau16}.}
\tablenotetext{b}{$N=(R_\mathrm{in}/D_{10})^2\cos\theta$, where $R_\mathrm{in}$
is the apparent inner disk radius in km, $D_{10}$ is the distance to the source
in 10\,kpc, and $\theta$ the angle of the disk \citep[$\theta=0$ is
face-on;][]{kub98}.}

\label{tab:modA}
\end{center}
\end{table*}

{\bf Model A:}

{\tt crabcorr*TBabs*(diskbb+relxilllpCp+xillverCp)}
\\

The {\tt crabcorr} model \citep{ste10}, is designed to standardize detector
responses to return the same normalizations and power-law slopes for the Crab,
assuming as a standard the \cite{too74} fit (i.e., $\Gamma=2.1$ and
$N=9.7$\,photons\,s$^{-1}$\,keV$^{-1}$).  The model spectrum of each dataset is
multiplied by a power law, applying both normalization and slope corrections.
We keep these quantities fixed for \nustar\ FPMA ($\Delta\Gamma=0$ and $N=1$).
Following Table~1 in \cite{ste10}, we fixed $\Delta\Gamma=-0.04$ for \swift\
XRT, while the rest are varied freely.

\begin{figure*}[ht!]
\centering
\includegraphics[width=0.5\linewidth]{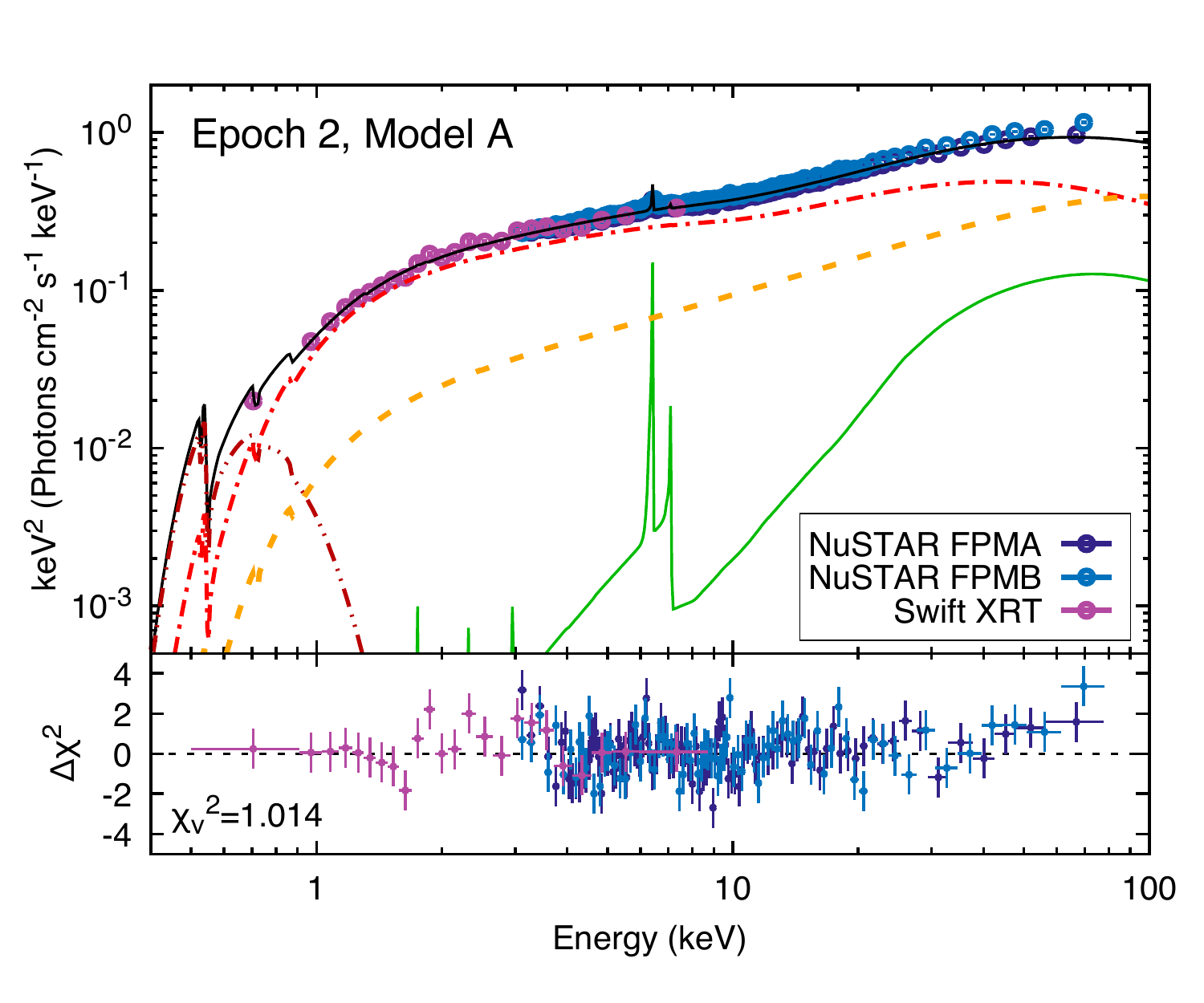}\includegraphics[width=0.5\linewidth]{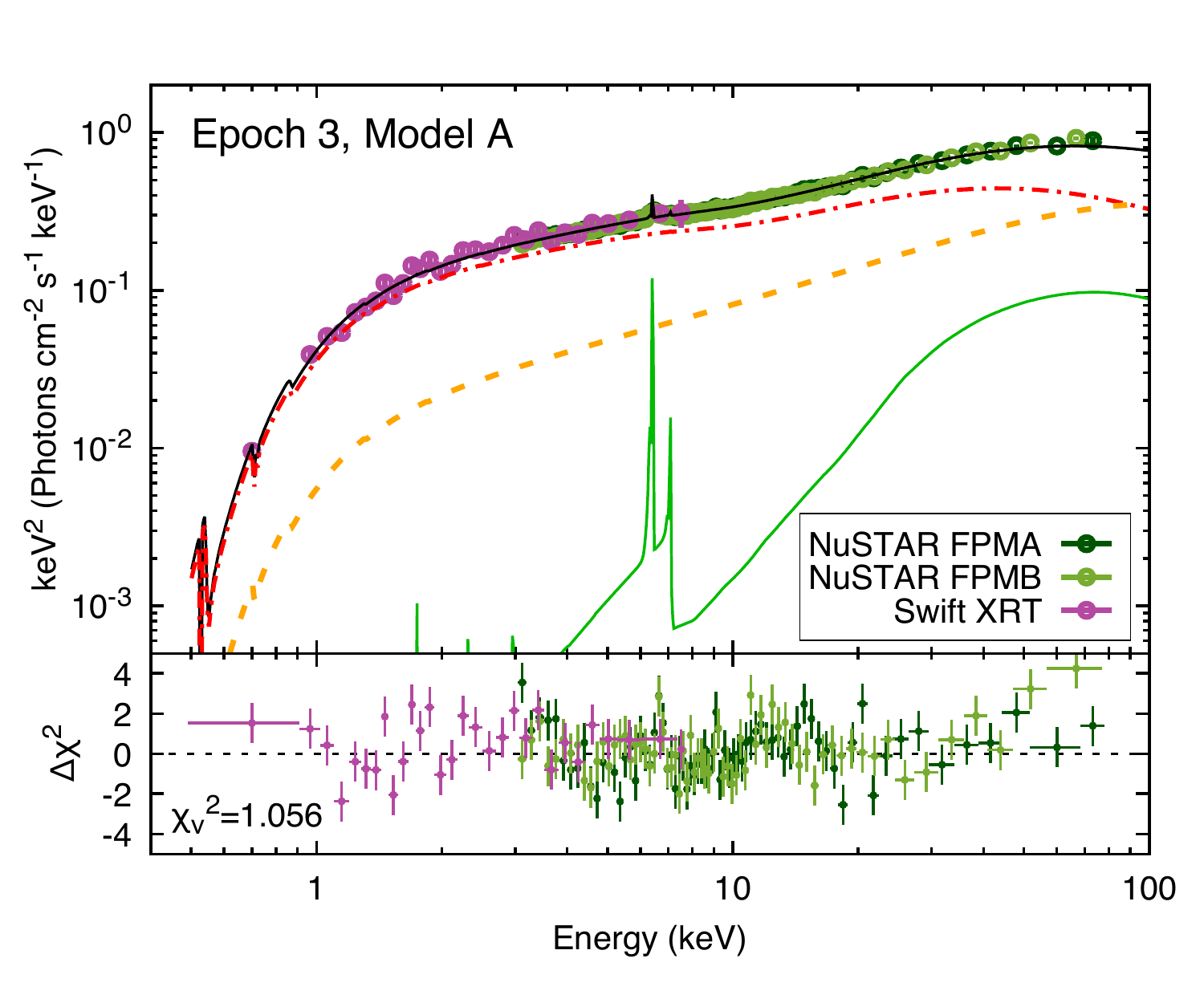}
\includegraphics[width=0.5\linewidth]{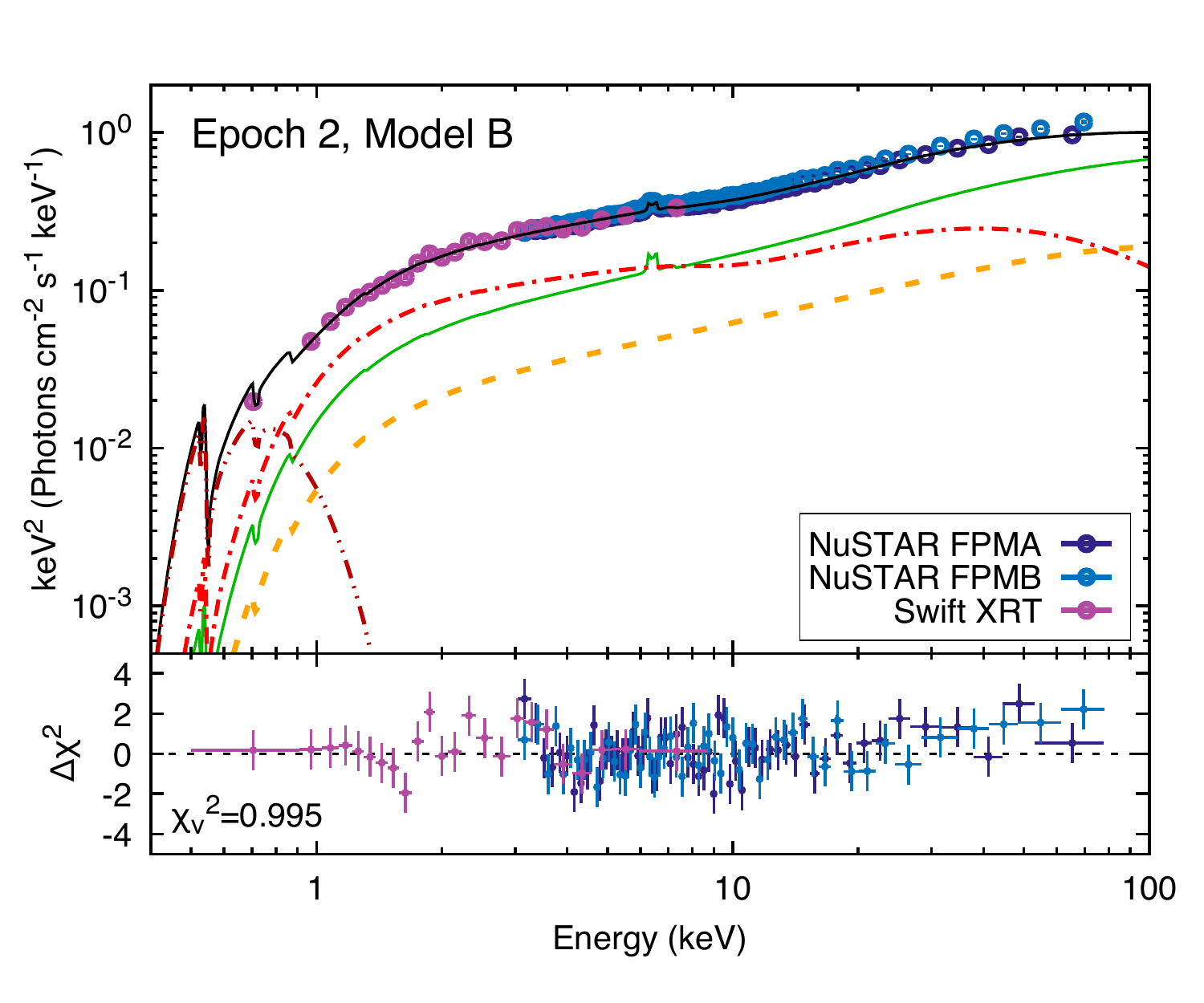}\includegraphics[width=0.5\linewidth]{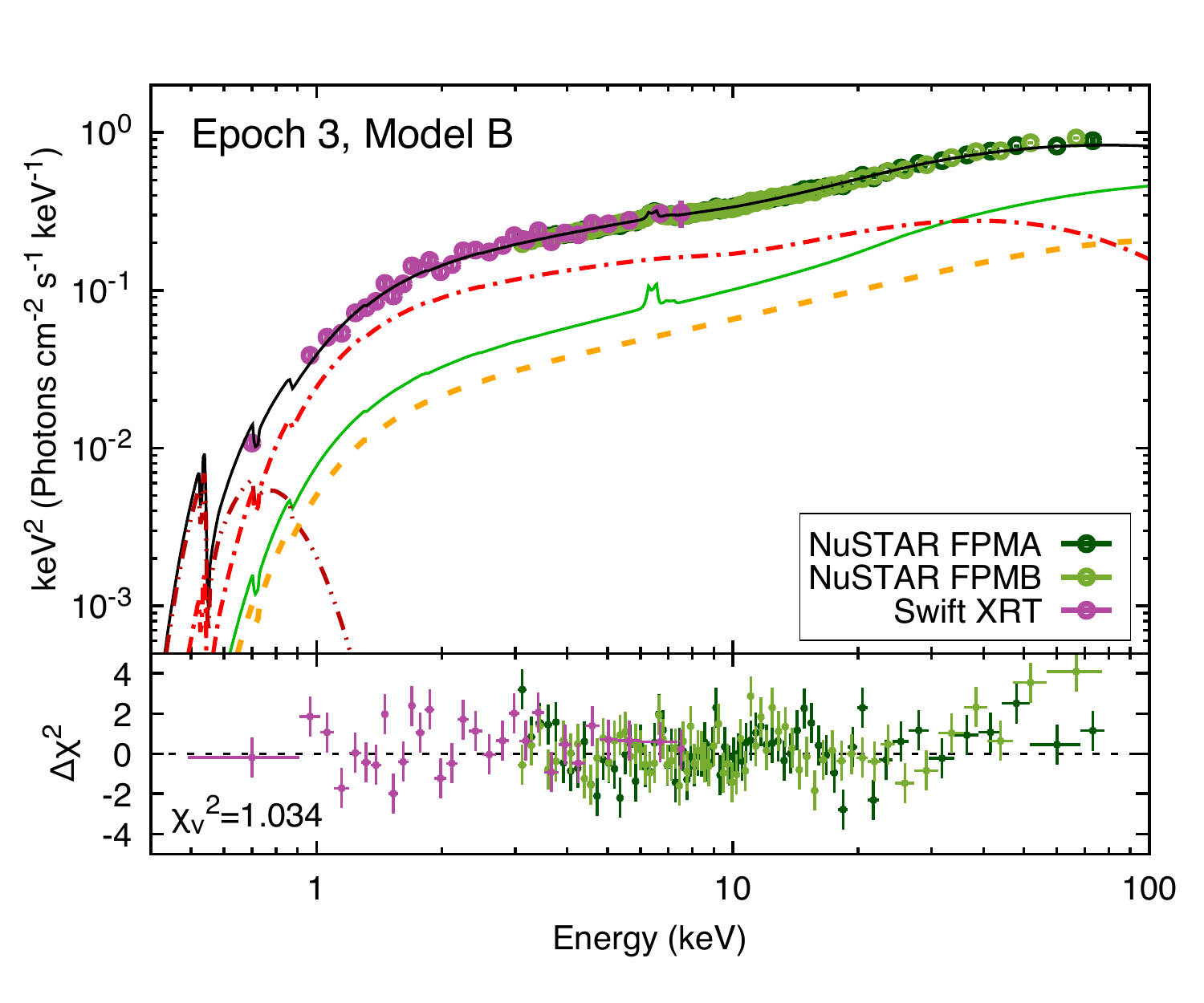}
\caption{
Analysis of the spectra from Epochs~2 and 3 (left and right), using relativistic reflection in a lamppost
geometry for the inner reflection and a non-relativistic, distant reflector (Model~A, top panels);
or a second lamppost located higher in the rotational axis (Model~B, bottom panels). The individual
model components are shown: total model (solid black), coronal emission (\nthComp; dashed orange), disk
emission ({\tt diskbb}; dot-dot dashed red), relativistic reflection from a lamppost (\relxilllpCp; dot-dashed red),
and distant reflection or second lamppost (\xillverCp\ or \relxilllpCp; solid green). The bottom sub-panels 
show the residuals of each fit, with the resulting $\chi^2_{\nu}$ indicated in each case.
}
\label{fig:fits}
\end{figure*}

We keep the spin parameter fixed at maximum ($a_*=0.998$) to allow for the maximum
possible disk truncation. We have found through several tests that the particular choice
of the spin value does not significantly influence the overall results. In general, lower
spin values provide a marginally worse fit (e.g., $\Delta\chi^2 \sim 4$ if $a_*=0.5$), and
both the inner radius and the height of the corona tend to decrease. Other parameters held fixed during the fit are the inclination,
disk outer radius, and the electron temperature. The reflection fraction of the \xillverCp\ component
is fixed to -1 such that no other continuum is added to the model. All the parameters associated with
this fit are summarized in Table~\ref{tab:modA}. The quality of the fit using Model~A
on Epochs~2 and 3 is equivalent to that in the last iteration of the progression
of models shown in the previous Section. The model requires that the primary source be
placed very close to the black hole ($h\sim 1.6$\,$R_\mathrm{Hor}$, where $R_\mathrm{Hor}$
is the radius of the event horizon), and for the
location of inner radius consistent with the ISCO. These parameters result in a
fairly large reflection fraction $R_\mathrm{f}\sim 8$, with more than $40$\% of the
primary photons lost into the black hole. The top panels of Figure~\ref{fig:fits} show
the different components of Model~A fitted to Epochs~2 and 3, together with their
respective residuals.

\def\fIIIEIInh{$7.0^{+0.7}_{-0.6}$}		 
\def\fIIIEIIh{$2.1^{+0.4}_{-0.5}$}		 
\def\fIIIEIIrin{$< 1.9$}			 
\def\fIIIEIIga{$1.50^{+0.04}_{-0.01}$}	 
\def\fIIIEIIlxi{$3.90^{+0.09}_{-0.09}$}		 
\def\fIIIEIIafe{$4.0^{+0.3}_{-0.5}$}		 
\def\fIIIEIIrxn{$1.2^{+0.8}_{-0.4}$}		 
\def\fIIIEIIhh{$503^{+160}_{-105}$}		 
\def\fIIIEIIrxnn{$4.5^{+0.9}_{-1.5}$}		 
\def\fIIIEIItin{$90^{+19}_{-17}$}		 
\def\fIIIEIIdin{$4^{+12}_{-4}$}		 
\def\fIIIEIIdga{$2.5^{+0.6}_{-0.6}$}		 
\def\fIIIEIIcrni{$0.97^{+0.01}_{-0.01}$}	 
\def\fIIIEIIcrnii{$0.93^{+0.02}_{-0.02}$}	 
\def\fIIIEIIchi{$1502.1$}			 
\def\fIIIEIIdof{$1510$}				 
\def\fIIIEIIred{$0.995$}			 
\def\fIIIEIIrf{$4.35$}                           
\def\fIIIEIIrs{$2.04$}                           
\def\fIIIEIIpf{$29.4$}                           

\def\fIIIEIIInh{$8^{+1}_{-1}$}		 
\def\fIIIEIIIh{$2.2^{+0.1}_{-0.3}$}		 
\def\fIIIEIIIrin{$< 4.5$}			 
\def\fIIIEIIIga{$1.50^{+0.01}_{-0.07}$}	 
\def\fIIIEIIIlxi{$4.03^{+0.02}_{-0.05}$}	 
\def\fIIIEIIIafe{$4.0^{+0.4}_{-0.8}$}	 
\def\fIIIEIIIrxn{$1.6^{+0.2}_{-0.3}$}		 
\def\fIIIEIIIhh{$> 765$}			 
\def\fIIIEIIIrxnn{$2.9^{+0.2}_{-0.4}$}		 
\def\fIIIEIIItin{$84^{+35}_{-31}$}	 
\def\fIIIEIIIdin{$< 184$}		 
\def\fIIIEIIIdga{$0.1^{+0.7}_{-0.7}$}		 
\def\fIIIEIIIcrni{$1.02^{+0.02}_{-0.01}$}	 
\def\fIIIEIIIcrnii{$0.87^{+0.02}_{-0.02}$}	 
\def\fIIIEIIIchi{$1532.7$}			 
\def\fIIIEIIIdof{$1483$}			 
\def\fIIIEIIIred{$1.034$}			 
\def\fIIIEIIIrf{$4.77$}                          
\def\fIIIEIIIrs{$2.14$}                          
\def\fIIIEIIIpf{$32.4$}                          

%
\begin{table*}
\begin{center}
\small
\caption{Best-fit parameters for Epochs~2 and 3 using the dual-lamppost (Model~B;
{\tt crabcorr*TBabs*(diskbb+relxilllpCp$_1$+relxilllpCp$_2$}). }
\begin{tabular}{lcrr}
\hline
Component & Parameter & Epoch~2 & Epoch~3 \\
\hline
{\tt relxilllpCp}$_{1,2}$ & $a_*$                                & \multicolumn{2}{c}{\fIIEIIa\ }    \\
{\tt relxilllpCp}$_{1,2}$ & $i$ (deg)                            & \multicolumn{2}{c}{\fIIEIIinc\ }  \\
{\tt relxilllpCp}$_{1,2}$ & $R_\mathrm{out}$ ($R_\mathrm{g})$    & \multicolumn{2}{c}{\fIIEIIrout\ } \\
{\tt relxilllpCp}$_{1,2}$ & $kT_\mathrm{e}$ (keV)                & \multicolumn{2}{c}{\fIIEIIkte\ }  \\
{\tt crabcorr}            & $\Delta\Gamma$ (XRT)                 & \multicolumn{2}{c}{\fIIEIIdg\ }   \\
\hline
{\tt TBabs}               & $N_\mathrm{H}$ ($10^{21}$ cm$^{-2}$) & \fIIIEIInh\   & \fIIIEIIInh\   \\
{\tt relxilllpCp}$_1$     & $h$ $(R_\mathrm{Hor})$               & \fIIIEIIh\    & \fIIIEIIIh\    \\
{\tt relxilllpCp}$_1$     & $R_\mathrm{in}$ ($R_\mathrm{ISCO})$  & \fIIIEIIrin\  & \fIIIEIIIrin\  \\
{\tt relxilllpCp}$_1$     & $\Gamma$                             & \fIIIEIIga\   & \fIIIEIIIga\   \\
{\tt relxilllpCp}$_1$     & $\log\xi$ (erg cm s$^{-1}$)          & \fIIIEIIlxi\  & \fIIIEIIIlxi\  \\
{\tt relxilllpCp}$_1$     & $A_\mathrm{Fe}$                      & \fIIIEIIafe\  & \fIIIEIIIafe\  \\
{\tt relxilllpCp}$_1$     & $N$ $(10^{-2})$\tablenotemark{a}     & \fIIIEIIrxn\  & \fIIIEIIIrxn\  \\
{\tt relxilllpCp}$_2$     & $h$ $(R_\mathrm{g})$                 & \fIIIEIIhh\   & \fIIIEIIIhh\   \\
{\tt relxilllpCp}$_2$     & $N$ $(10^{-3})$\tablenotemark{a}     & \fIIIEIIrxnn\ & \fIIIEIIIrxnn\ \\
{\tt diskbb}              & $kT_\mathrm{in}$ (eV)                & \fIIIEIItin\  & \fIIIEIIItin\ \\
{\tt diskbb}              & $N$ $(10^{6})$\tablenotemark{b}      & \fIIIEIIdin\  & \fIIIEIIIdin\ \\
{\tt crabcorr}            & $\Delta\Gamma$ ($10^{-2}$, FPMB)     & \fIIIEIIdga\   & \fIIIEIIIdga\ \\
{\tt crabcorr}            & $N$ (FPMB)                           & \fIIIEIIcrni\  & \fIIIEIIIcrni\ \\
{\tt crabcorr}            & $N$ (XRT)                            & \fIIIEIIcrnii\ & \fIIIEIIIcrnii\ \\
\hline
Reflection Fraction       & $R_\mathrm{f}$                       & \fIIIEIIrf\   & \fIIIEIIIrf\  \\
Reflection Strength       & $R_\mathrm{s}$                       & \fIIIEIIrs\   & \fIIIEIIIrs\  \\
Photons Lost in BH        & $N_\mathrm{ph}$ (\%)                 & \fIIIEIIpf\   & \fIIIEIIIpf\  \\
\hline
\multicolumn{2}{c}{$\chi^2$}                                     & \fIIIEIIchi\  & \fIIIEIIIchi\ \\
\multicolumn{2}{c}{$\nu$}                                        & \fIIIEIIdof\  & \fIIIEIIIdof\ \\
\multicolumn{2}{c}{$\chi_{\nu}^2$}                               & \fIIIEIIred\  & \fIIIEIIIred\ \\
\hline
\end{tabular}
\tablenotetext{a}{The normalization of \xillver\ is defined such that for a source spectrum
with flux $F_\mathrm{x}(E)$ incident on a disk with density $n_e$, then
$\int_{0.1\mathrm{keV}}^{1\mathrm{MeV}}{F_\mathrm{x}(E)dE}=10^{20}\frac{n_e\xi}{4\pi}$, where
$\xi$ is the ionization parameter. For the \relxill\ models the definition is identical, although
the observed flux differs due to the relativistic effects \citep[see][]{dau16}.}
\tablenotetext{b}{$N=(R_\mathrm{in}/D_{10})^2\cos\theta$, where $R_\mathrm{in}$
is the apparent inner disk radius in km, $D_{10}$ is the distance to the source
in 10\,kpc, and $\theta$ the angle of the disk \citep[$\theta=0$ is
face-on;][]{kub98}.}
\label{tab:modB}
\end{center}
\end{table*}

Despite the success of Model~A, careful inspection of Figure~\ref{fig:fits} shows some structure in the
residuals near the Fe K region, which suggests that the narrow component of the iron
emission is perhaps more complex than the one produced by a distant reflector. We
have thus tried an additional fit, in which the \xillverCp\ component is replaced
by a second lamppost, one situated at a larger height than the first one. We refer
to this as Model~B, written as
\\

{\bf Model B:}

{\tt crabcorr*TBabs*(diskbb+relxilllpCp}$_1${\tt +relxilllpCp}$_2${\tt )}
\\

In this Model~B, all the parameters of the second lamppost are tied to the first one,
except for the height and the normalization. As in the case of the distant reflection,
we set the reflection fraction of the second lamppost to -1, such that only one continuum
is included by the model (via the first lamppost). 
Furthermore, given the poor constraint
on the inner radius, and to ensure that the second lamppost only provides reflection
from farther away in the disk, we assume that \rin$_2$ is equal to the height of the
lamppost (this implies that the second lamppost only produces reflection in a disk
with a much larger inner radius). With only one extra d.o.f., the dual lamppost (Model~B) provides
significantly better fit statistics ($\Delta\chi^2 = 30$ for Epoch~2, and $\Delta\chi^2 = 37$
for Epoch~3). The best fit parameters are summarized in Table~\ref{tab:modB}, while
the model components and residuals are shown the bottom panels of Figure~\ref{fig:fits}.

%

\section{Discussion}\label{sec:disc}

The spectral fits detailed in the previous Sections reveal several
important aspects of the evolution of \gx339\ in the early stages of its
frequent outbursts. We find that even at very low accretion rates, such as
those observed during Epochs~1 and 4 ($\sim 0.2$\% \ledd), the spectrum observed
by \nustar\ shows clear hallmarks of X-ray reprocessing in an optically-thick
medium, likely the accretion disk. These include Fe K emission near 6.4\,keV,
a smeared Fe K-edge at $\sim7-10$\,keV, and possibly a Compton hump at higher
energies (20--40\,keV).  Given the limited signal of these low-luminosity
observations, we cannot confidently determine the width of the Fe K emission, 
nor constrain the disk inner radius. Interestingly, the equivalent width of 
the Fe K emission measured in Epochs~1 and 4 (\fIeqwEI\,eV and \fIeqwEIV\,eV) 
are larger than in Epochs~2 and 3 (\fIeqwEII\,eV and \fIeqwEIII\,eV),
contrary to the expectation of a stronger emission if the inner accretion disk
moves inward when the luminosity increases. This is likely due to the fact that
the {\tt smedge} component can be unphysically large, resulting in strong degeneracies.
However, the EW measured in Epoch~1
agrees well with the value of $73^{+18}_{-14}$\,eV previously measured by 
\cite{tom09} with \suzaku\ data taken at a similar luminosity level.

Testing different types of reflection models on the highest luminosity
observations (Epochs~2 and 3), we find that both relativistic (broad)
and non-relativistic (narrow) components are strongly required to fit
the data. Our two main fits are then based on these findings. Model~A
includes relativistic reflection from the inner accretion disk invoking
a lamppost geometry, as well as 
a non-relativistic reflection component assumed to be produced much 
farther, possibly in the outer disk or the surface of the companion star.
Meanwhile, Model~B replaces the non-relativistic reflection for a second
lamppost, which is placed at a much larger height, and thus 
preferentially illuminates larger radii in the disk,
producing a narrower Fe K
emission than the lamppost located close to the black hole.

While both Models~A and B reproduce the data well, Model~B yields better
fit statistics, with \redcsq=0.995 for Epoch~2, which represents an improvement
of $\Delta\chi^2=30$ for only one extra d.o.f. Fits to Epoch~3 also resulted
in a preference for Model~B (\redcsq=1.034; $\Delta\chi^2=37$). Most of this
improvement comes from a better fit of the narrow component of the Fe K emission,
which suggest that this narrow component could in fact originate in a
region of the disk 
at sufficiently large radii such that relativistic effects are largely negligible,
but close enough to still be affected by the rotational motion of the disk.
Similar complexity has been seen in the narrow Fe K component in the black hole
binary candidate MAXI~J1535$-$571, independently by both \nustar\ \citep{xuy18},
and \nicer\ \citep{mil18}, and also possibly in the black hole binary
candidate MAXI~J1820+070 (Buisson et al. in prep).

Although the lamppost is a highly idealized geometry, it provides a good
description of the observed X-ray spectrum, and allows for the determination
of several key parameters, such as how close to the black hole the primary
source is located, and the exact relative strength of the direct and reflected
components (i.e., the reflection fraction). Furthermore, the dual lamppost description
(Model~B) is not only superior in terms of the fit statistics, but also
in the overall quantities recovered by the fit. The height of the source
is slightly higher in Model~B, which implies a more reasonable reflection
fraction ($R_\mathrm{f}\sim4-5$ vs $8-9$ for Model~A), which reduces the
fraction of photons lost into the black hole from $\sim40$\% to $\sim30$\%,
relaxing the high luminosity implied for the primary source. We notice
that the photon index increases from $\Gamma=1.37$ to $1.5$ between Models~A
and B, with the latter value being more consistent with values reported
in previous analysis \citep[e.g.,][]{fue15,gar15,wan18}.

The ionization parameter, which is proportional to the ratio of the incident
flux to the gas density ($\xi = 4\pi F_x/n_e$), is found to be significantly
larger than in previous studies at similar accretion states \citep[e.g.,][]{fue15,
gar15,wan18}. This is true for both Models~A and B, and in both Epochs~2 and 3. 
Taking the case of Epoch~2, the 2--10\,keV observed flux is comparable to that
in Obs.~1 (2015) of \cite{wan18}, who reported $\log\xi\sim3.3$,
while we find $\log\xi\sim4$. Assuming that the fluxes in the two observations
are in fact identical, the difference can only be attributed to a change in the
density of the disk's atmosphere by a factor of $\sim 5$. This difference does 
not seem implausible, given the intrinsic turbulent nature of accretion disks
\citep[e.g.,][]{kad18}, and how strongly magnetic fields can affect the surface
density in the accretion disk \citep[e.g.,][]{jiayf19}.

In the case of Epoch~3, we found that when using Model~A the thermal disk
component ({\tt diskbb}) is not statistically required, as it produces
virtually no change in the goodness of the fit, and the model parameters are
unconstrained. The situation is different when using Model~B, in which case we
found an improvement of $\Delta\chi^2=8$ when adding the thermal disk
component.  This is likely due to the smaller reflection fraction, which
lowers the reflected continuum and makes the disk emission more obvious in the
fit.  Nevertheless, we notice that this spectral feature is not very prominent,
and its origin cannot be fully determined. For example, \cite{gar16b} showed
that when the density in the reflector is over $\sim 10^{17}$\,cm$^{-3}$ there
is an enhancement of reflected flux at soft energies. As mentioned above,
the relatively large ionization parameter indicates the possibility of an also
large density in the accretion disk, which could then explain the soft excess
observed in the spectra. As we discuss next, these high-density effects can
also have important consequences in the abundance of iron derived form
reflection spectroscopy. 

The iron abundance for Model~A is constrained at more than 6 times its
Solar value, while for Model~B (dual lamppost) it is found to be $\sim4$ times solar.
Both of these results are consistent with our previous analysis of \rxte\ PCA
\citep{gar15} and \nustar\ data \citep{wan18}. However, \cite{jia19} has recently demonstrated
that by implementing new reflection models calculated at high densities 
the requirement for the disk thermal emission vanishes, and the recovered
iron abundance is more consistent with the Solar value. Since these high
density models are still under development, and given that all the other
parameters remain unchanged in the analysis of \cite{jia19}, we deferred
their application to these data for a future publication.

\begin{figure}[ht!]
\centering
\includegraphics[width=\linewidth]{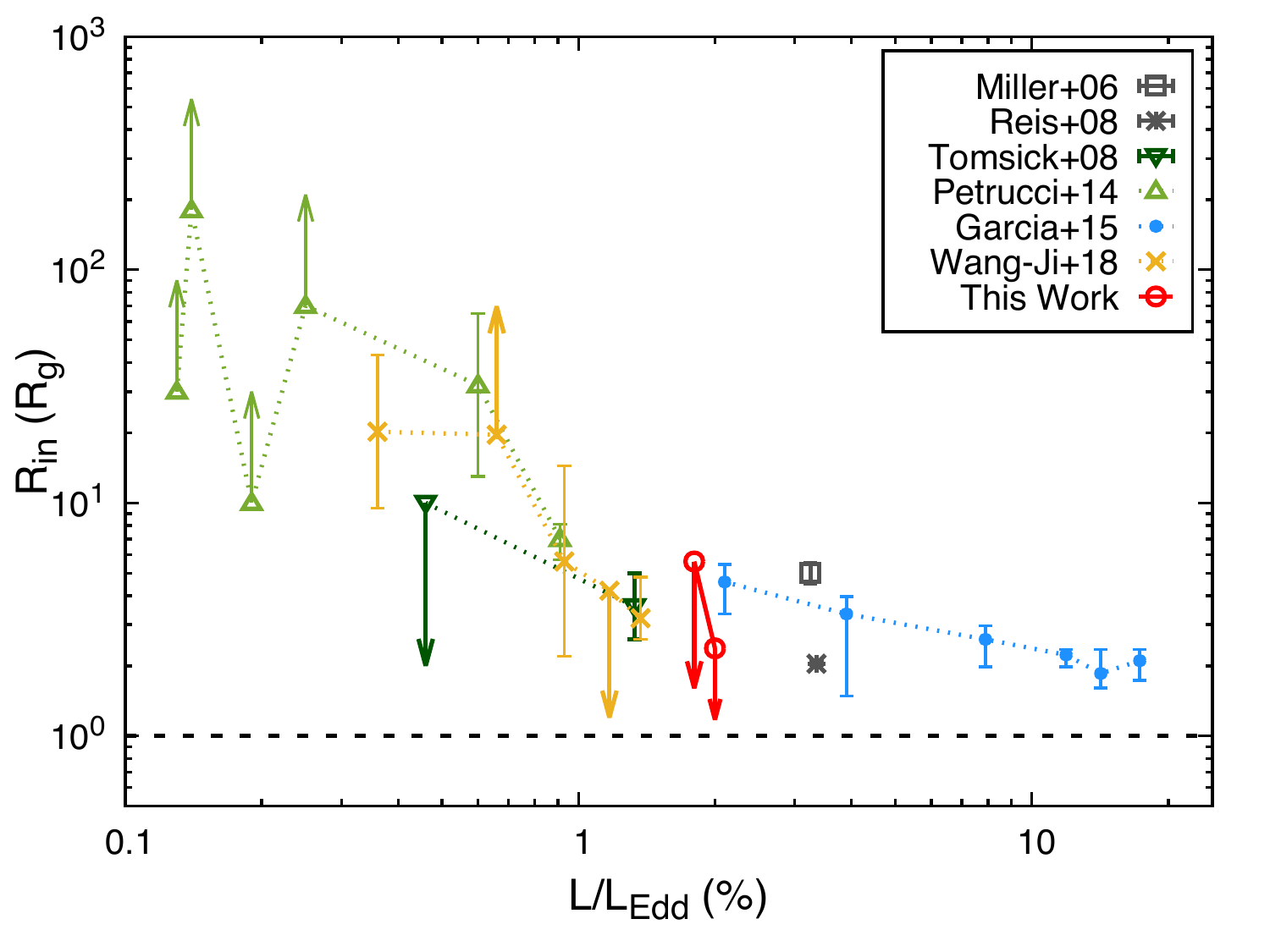}
\caption{Comparison of the inner radius values derived from reflection spectroscopy
as a function of the Eddington-scaled luminosity for the hard state of \gx339. Reported
values include those from analysis of \xmm\ MOS \citep{mil06,rei08}, \suzaku\ \citep{tom08,pet14},
\rxte\ PCA \citep{gar15}, and \swift+\nustar\ \citep{wan18}. Our values derived from the
2017 \swift+\nustar\ data are in good agreement with the observed correlation between
\rin\ and the source luminosity.
}
\label{fig:rin}
\end{figure}

It is important to notice that both of these models constrain the inner radius to be
very close to the ISCO radius, specifically \rin$<1.2$\,\risco\ for
Model~A (distant reflector), and \rin$=2-5$\,\risco\ for Model~B (dual lamppost).
Figure~\ref{fig:rin} shows a comparison our results with several other inner radius 
measurements of \gx339\ with reflection spectroscopy. The values of \rin\
(in units of gravitational radius $R_g = GM/c^2$) are plotted as a function
of the Eddington scaled luminosity \ledd$=1.25\times 10^{39}$\,erg\,s$^{-1}$
\citep[assuming a distance of $D=8$\,kpc and a black hole mass of $M=10$\,\msun;][]{zdz04}.
The results from our analysis of the 2017
failed outburst data are in good agreement with the overall trend of a 
decreasing inner radius with increasing luminosity in the hard state.
Furthermore, the trend observed in Figure~\ref{fig:rin} indicates that the
inner accretion disk appears to be relatively close to the ISCO radius early
on in the outburst, reaching few times \risco\ at $\sim1$\% \ledd.
Interestingly, this trend appears to be independent of whether the source was observed
during the rise or the decay of the outburst, and on whether the outburst
itself was full (i.e., the source went through state transitions), or a 
failed one (such as the one analyzed here).

%

\section{Conclusions}\label{sec:concl}

The black hole binary system \gx339\ goes into outburst regularly, with a full outburst
typically observed every 2--3 years \citep{tet16}. In this paper we have presented
a reflection spectroscopy analysis of the X-ray spectrum of \gx339\ as observed by
\nustar\ and \swift\ during the 2017 failed outburst. We triggered 3 observations
on the rise in the hard state. The source reached $\sim 50$\,mCrab (2--10\,keV),
before observations were restricted due to Sun constraint during December 2017. 
One more observation done at the end of January 2018 showed that the source did 
not transition to the soft state, and it was already on the decay to quiescence.

The \nustar\ spectra from our observations show clear signatures of X-ray reflection
from an optically-thick accretion disk (i.e., Fe K emission, K-edge, and Compton hump).
Detailed spectroscopic analysis for the two brightest observations revealed the need
for a relativistically broadened reflection component, as well as a narrow and likely
more distant reflection component; in addition to a non-thermal (power-law like) continuum,
and weak thermal disk emission. Our analysis focuses on two main fits, both including
a relativistic reflection model using the lamppost geometry.

The dual-lamppost description is preferred over the single lamppost plus a
simple distant reflector, based on a relatively significant improvement of the
fit statistics of the two highest S/N spectra in our sample. The second
lamppost, which is situated at a much larger height than the first, provides a
better fit to the narrow component of the Fe K emission.  This picture is
consistent with a corona that has a certain vertical dimension and is
capable of illuminating the outer regions of the accretion disk.
Meanwhile, we do not
detect any changes in the coronal emission, as the slope of the non-thermal
continuum is consistent among the four observations analyzed here, and no
high-energy cutoff can be detected (and thus fixed in our fits). 

The results from these fits are in good agreement with our recent analysis done
with the same reflection models on \nustar\ and \swift\ data from the previous
2015 and 2013 outbursts \citep{wan18}, as well as with those derived from the
analysis of \rxte\ observations \citep{gar15}. Particularly, we find that the
inner radius of the accretion disk is close to the ISCO in the two models adopted
here. Most importantly, our measurements are fully consistent with the
observed trend of \rin\ decreasing as the luminosity in the hard state increases.
The present analysis shows that the inner disk must be close to the ISCO early
on during the outburst, and that it is typically within a few ISCO radii for
luminosities of $\sim1$\% Eddington. Crucially, this trend has been constructed with data
provided by different observatories such as \suzaku, \swift, \rxte, and \nustar;
with observations taken during full and failed outbursts, and during both the rise and decay
phases of the hard state evolution. This implies that the state transitions between
hard and soft states are unlikely to be triggered by changes in the location of
the inner accretion disk, but rather by other mechanisms. Future and continuous
monitoring of this and other black hole transient sources will provide valuable
new insights into the accretion physics of these systems.

%
%
%
\acknowledgments 
We thank the anonymous referee for the careful revision of this paper. We also thank
Erin Kara and Didier Barret for enlightening discussions on the implications of the dual
lamppost model.
J.A.G. acknowledges support from NASA grant NNX17AJ65G and from the Alexander
von Humboldt Foundation.  R.M.T.C. has been supported by NASA ADAP grant
80NSSC177K0515.
V.G. is supported through the Margarete von Wrangell fellowship by the 
ESF and the Ministry of Science, Research and the Arts Baden-W\"urttemberg.
N.S. would like to acknowledge the support from DST-INSPIRE
and Caltech SURF-2017 fellowships.  This work was partially supported under
NASA contract No. NNG08FD60C and made use of data from the {\it NuSTAR}
mission, a project led by the California Institute of Technology, managed by
the Jet Propulsion Laboratory, and funded by the National Aeronautics and Space
Administration. We thank the {\it NuSTAR} Operations, Software, and Calibration
teams for support with the execution and analysis of these observations. This
research has made use of the {\it NuSTAR} Data Analysis Software (NuSTARDAS),
jointly developed by the ASI Science Data Center (ASDC, Italy) and the
California Institute of Technology (USA).

\vspace{5mm}
\facilities{\it NuSTAR, Swift}
\software{{\sc xspec} \citep[v12.10.0c;][]{arn96},
{\sc xillver} \citep{gar10,gar13a}, {\sc relxill}
\citep[v1.2.0;][]{gar14a,dau14}, {\sc nustradas} (v1.6.0)}

%
%
%
%
\bibliographystyle{aasjournal}
\bibliography{my-references}
%
%
%
%
\end{document}